\documentclass[10pt,letterpaper]{article}
\usepackage[top=0.85in,left=2.75in,footskip=0.75in]{geometry}

\usepackage{amsmath,amssymb}

\usepackage{changepage}

\usepackage{textcomp,marvosym}

\usepackage{cite}

\usepackage{nameref,hyperref}

\usepackage[right]{lineno}

\usepackage[nopatch=eqnum]{microtype}
\DisableLigatures[f]{encoding = *, family = * }

\usepackage[table]{xcolor}

\usepackage{array}

\newcolumntype{+}{!{\vrule width 2pt}}

\newlength\savedwidth

\raggedright
\usepackage[aboveskip=1pt,labelfont=bf,labelsep=period,justification=raggedright,singlelinecheck=off]{caption}
\renewcommand{\figurename}{Fig}

\makeatletter
\renewcommand{\@biblabel}[1]{\quad#1.}
\makeatother

\usepackage{lastpage,fancyhdr,graphicx}
\usepackage{epstopdf}
\fancyheadoffset[L]{2.25in}
\fancyfootoffset[L]{2.25in}
\begin{document}
\vspace*{0.2in}

\begin{flushleft}
{\Large
\textbf\newline{From atomic to global connectivity in the structure of the SARS-CoV2-Human ACE2 receptor complex} 
}
\newline
\\
Varsha Subramanyan\textsuperscript{1},
Arinnia Anto\textsuperscript{2,3\ddag},
Moitrayee Bhattacharyya\textsuperscript{4},
Smitha Vishveshwara\textsuperscript{5},
Saraswathi Vishveshwara\textsuperscript{2*}
\\
\bigskip
\textbf{1} Theoretical Division, Los Alamos National Laboratory, Los Alamos, New Mexico, United States of America
\\
\textbf{2} Molecular Biophysics Unit, Indian Institute of Science, Bangalore, India
\\
\textbf{3} University of Regina, Regina, Canada
\\
\textbf{4} Department of Pharmacology, Yale University, New Haven, Connecticut, United States of America
\\
\textbf{5} Department of Physics, University of Illinois at Urbana-Champaign, Urbana, Illinois, United States of America
\bigskip

%
%

*Corresponding author: saraswathi@iisc.ac.in


\ddag Current Affiliation

\end{flushleft}
\section*{Abstract}
We investigate connectivity properties of the SARS-CoV2 spike protein-human ACE2-receptor
complex employing a protein side chain-based network method that allows us to span a range from
atomic to global protein scales. We analyze network topology in terms of clusters and cliques
obtained from averaging over snapshots of MD simulations (from D.E. Shaw Research). We
demonstrate that SARS-CoV2 forms a more dominant, robust connection with the ACE2-receptor as
compared to the less virulent SARS-CoV1. Globally, this stronger connectivity is reflected by our
percolation analysis
where the interface cluster for the SARS-CoV2-ACE2 complex persists when restricted to  stronger bonds, as
compared to the SARS-CoV1- ACE2 complex. At the atomic level, the way in which network cliques form at the interface also reflects stronger connectivity for SARS-CoV2.  We pinpoint key functional residues in SARS-CoV2 that
play important roles in establishing this higher connectivity. Thus,
our studies provide an objective method to map spatial connectivity of atomic level non-covalent
interactions to global connectivity between any two amino acids in the complex. We also analyze
specific snapshots of the
MD simulations to highlight prominent variations in network topology that explore diverse
conformational landscapes. Finally, we demonstrate that most mutations that occur in the
SARSCoV2 spike protein in variants of concern/interest (including the currently active JN.1 and their subvariants) have
been observed at the interface with the ACE2 receptor. Our analyses highlight the importance of
interface interactions and provide a rationale for designing receptor-like peptides and proteins to
combat immunity-escaping variants.

\section*{Author summary}
As the worldwide grip of the COVID-19 virus continues in its ever-mutating forms, progress in
combating it would benefit from alternate perspectives of its protein structure and the complex it forms
with the human receptor. Here, we present a protein side-chain network characterization and
percolation analysis of the complex, its connectivity, and its interface. We highlight the stronger
interaction between the spike protein of COVID-19 and the human receptor in comparison to earlier coronaviruses. We
show that the connectivity map hosts a wealth of information regarding mutations and the global
structural changes brought about by local variations.

\section*{Introduction}
The outbreak of the global COVID-19 pandemic was an unprecedented humanitarian crisis demanding large scale cooperation between governments, scientists and healthcare workers for its alleviation. The Nobel Prizes for Chemistry in 2022 and Medicine in 2023 stand as a testament to the scientific progress made during this period in studying the virus, modeling its spread, understanding its effects and importantly, the development and effective distribution of vaccines. It has encompassed experimental, theoretical and computational approaches in various disciplines, including clinical research,  immunology, structural and molecular biology, simulations, structural modeling, and network-based analyses. A large body of literature has been generously made available to the public, a few examples of which are referenced here \cite{rf1,rf2,rf3,rf4,rf4a,rf5,rf6,rf7,rf8,rf9,rf10,rf11,rf12,rf13,rf14,rf15,rf16,Nussinov2022,Levitt,Atilgan2021,DiPaola2021}.
However, despite the availability of  vaccines, the spread of infections and the virulence of SARS-CoV2 variants, continues to be a cause for concern. Mutations in the SARS-CoV2 spike gene have altered protein binding efficiency and immunogenicity with subsequent lineages leading to variants with higher virulence and/or transmissibility. 
The evolutionary trajectory of the virus depends on the complex interplay of several factors including the status of vaccination, history of infection, complications in immuno-compromised patients etc. Notably, mutations in variants generally do not occur at single points, instead taking place in multiple locations of the spike protein. A complete understanding of external factors and such mutations on the functions of the virus- binding efficacy, immunogenicity, and evasive strategies is yet to emerge. 

A more holistic approach that at once incorporates atomic level interactions and provides global levels of abstraction is required. In our work, we take such an approach,  particularly focusing on the receptor binding domain (RBD) of the SARS-COV2 spike protein with the human ACE2 receptor. {\color{black} An outlook that we find inspiring in this context draws focus to biomolecules as poised at the confluence of simplicity and complexity\cite{Frauenfelder} wherein the building blocks together form an entity that can process information, adapt to the environment, and emerge as an 'intelligent' collective whole\cite{Tripathi2025-rf}. Our network approach reflects these perspectives in bridging between quantum chemistry features at the atomic level and the mesoscopic level of an entire protein. At the heart, we utilize network theory as the link in understanding Biology from atomic level physical approaches to the real-life functioning of a protein as a complex, cohesive whole. }

Here, we have investigated the 3D structures of Human ACE2 receptor bound to SARS-CoV2 and SARS-CoV1 from a global non-covalent connectivity perspective, by explicitly considering the sidechain atomic details and their interactions. We have analysed 10$\mu$s simulation trajectories \cite{DEShaw} to characterize the interaction of the receptor binding domain (RBD) of the spike proteins with that of the ACE2 receptor. To this end, we construct a protein sidechain network (PScN) from the 3D protein complexes, as laid out in previous works, with relevant details presented in the Methods section. Compared to conventional methods, the metrics of the PScN - interface clusters, hubs, cliques and so on - offer crucial clues in understanding both the global connectivity of the network as well as local connectivity between key regions.  It is also notable that such network metrics are dependent on edge interactions which are dynamic and fluctuate over time in equilibrium simulations \cite{rf8}.  Our earlier works \cite{Frontiers,jcim,Bhattacharyya2016-qt,Bhat,KANNAN1999441} extensively address this issue through graph spectral analysis alongside concepts borrowed from statistical  physics. In particular, we demonstrate that these networks show percolation-like behaviour with respect to edge strength when the networks are dynamically averaged over the length of the simulation.   

The importance of protein side chain network (PScN) studies can be appreciated by a brief description of the sequence of the spike protein of Covid-19 (belonging to the family of $\beta$-coronaviruses), indicating the structural and dynamical regions, relevant to its function. This is captured in two pictorial forms from our previous studies (Ref. \cite{Frontiers}), as shown in \ref{S1 Fig} Fig and \ref{S2 Fig} Fig (Figs 2 and 1, reproduced from the Ref \cite{Frontiers}). The homo-trimeric complex of SARS-CoV2 is crucial to the viral entry into the host cell, angiotensin-converting-enzyme-2 receptor (ACE2-Receptor). It exists in closed and open states, respectively in the inactive and the active forms of the virus. The differences of the two states in terms of allosteric effect and reorganization of the higher order connectivity parameters like cliques, communities, and clusters are presented in Ref \cite{Frontiers}.

We apply these network methods to compare the global connectivity of both SARS-CoV1 and SARS-CoV2 bound to the human ACE2 receptor which forms a protein complex. Our results show that globally, the largest cluster in the network is at the interface between the spike protein and ACE2 receptor in both cases, with the SARS-CoV2 complex forming a more robust and dominant cluster in comparison to SARS-CoV1. At the atomic level, the interface cliques and communities reveal a more tightly held complex for SARS-CoV2. We also present analysis of certain specific snapshots chosen from the simulation trajectory to highlight fluctuations in time which are not captured by the dynamical averages, thus demonstrating the diversity of conformational landscapes explored during the simulation. Further, we also locate the mutated residues in key variants of concern/interest, which notably occur at the interface between the spike protein and the human receptor, further indicating its crucial role in global connectivity. Our analysis thus provides an atomistic map of the entire protein complex, based on optimal connectivity between spatially proximal amino acid residues. Such a map can act as a bridge to connect the local features with those observed or inferred from the global properties, thus providing a possible mechanistic explanation for the higher transmitability/virulence of some variants over others. Therefore, our methods provide a basis to understand structure-function relationships and assess the outcome and design of drugs and antibodies in a rational manner.

\section*{Results and Analyses}

\begin{figure}
    \centering
    \includegraphics[width=\textwidth]{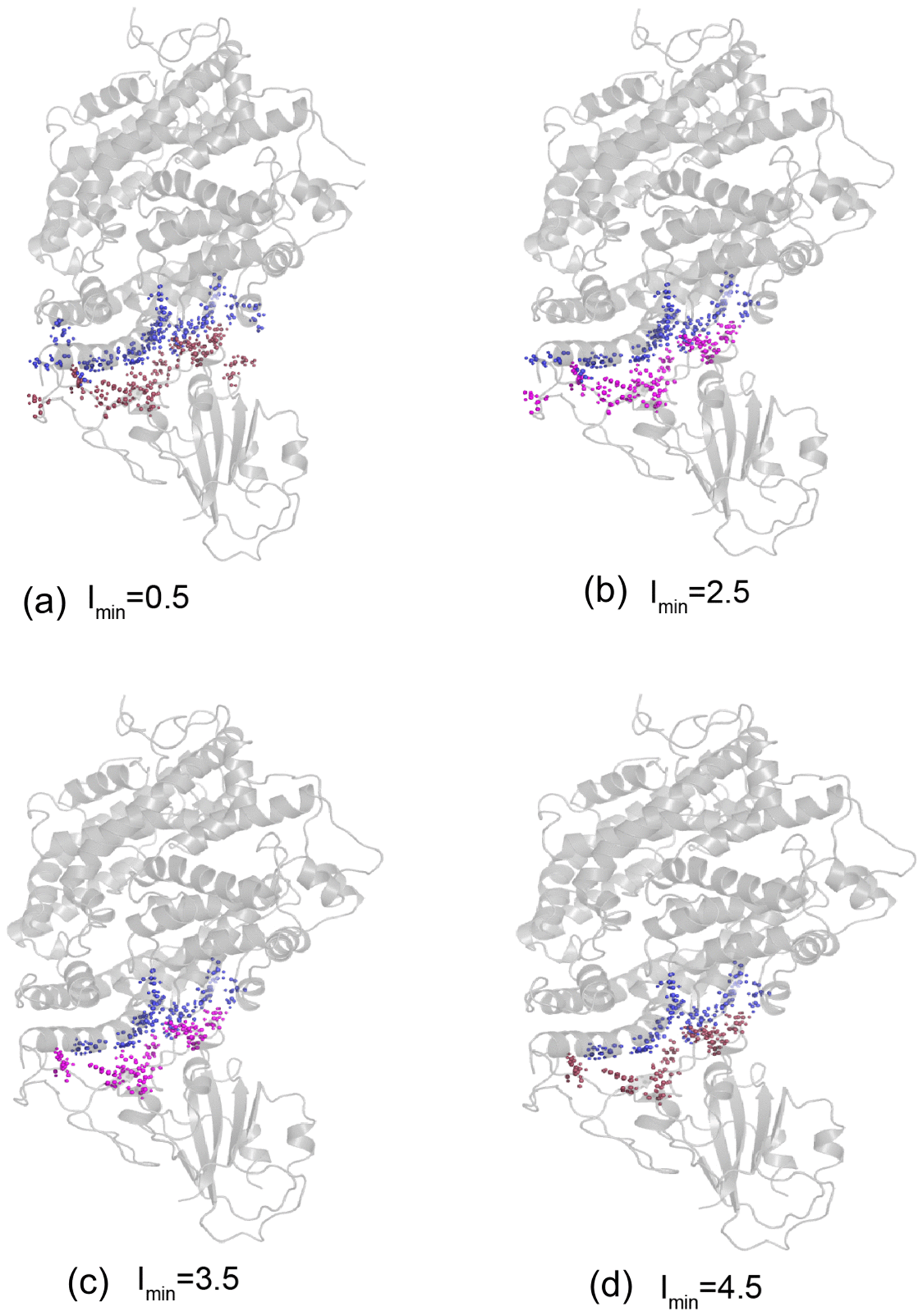}
    \caption{{\bf SARS-CoV1 Interface Clusters:} The PYMOL representation of the interface clusters of SARS-CoV1 complex at various $I_{min}$ at 50$\%$ dynamical stability are presented here. The nodes corresponding to ACE2 receptor residues are represented in blue, while virus residues are represented in pink. The breaking up of the largest interface cluster for increasing values of $I_{min}$ is indicative of the percolation-like behaviour presented in subsequent figures. The complex shows a reduction in the size of the interface cluster at $I_{min}$ $\sim$ 2.5 as corroborated in these representations.}
    \label{Fig 1}
\end{figure}

\begin{figure}
    \centering
    \includegraphics[width=\textwidth]{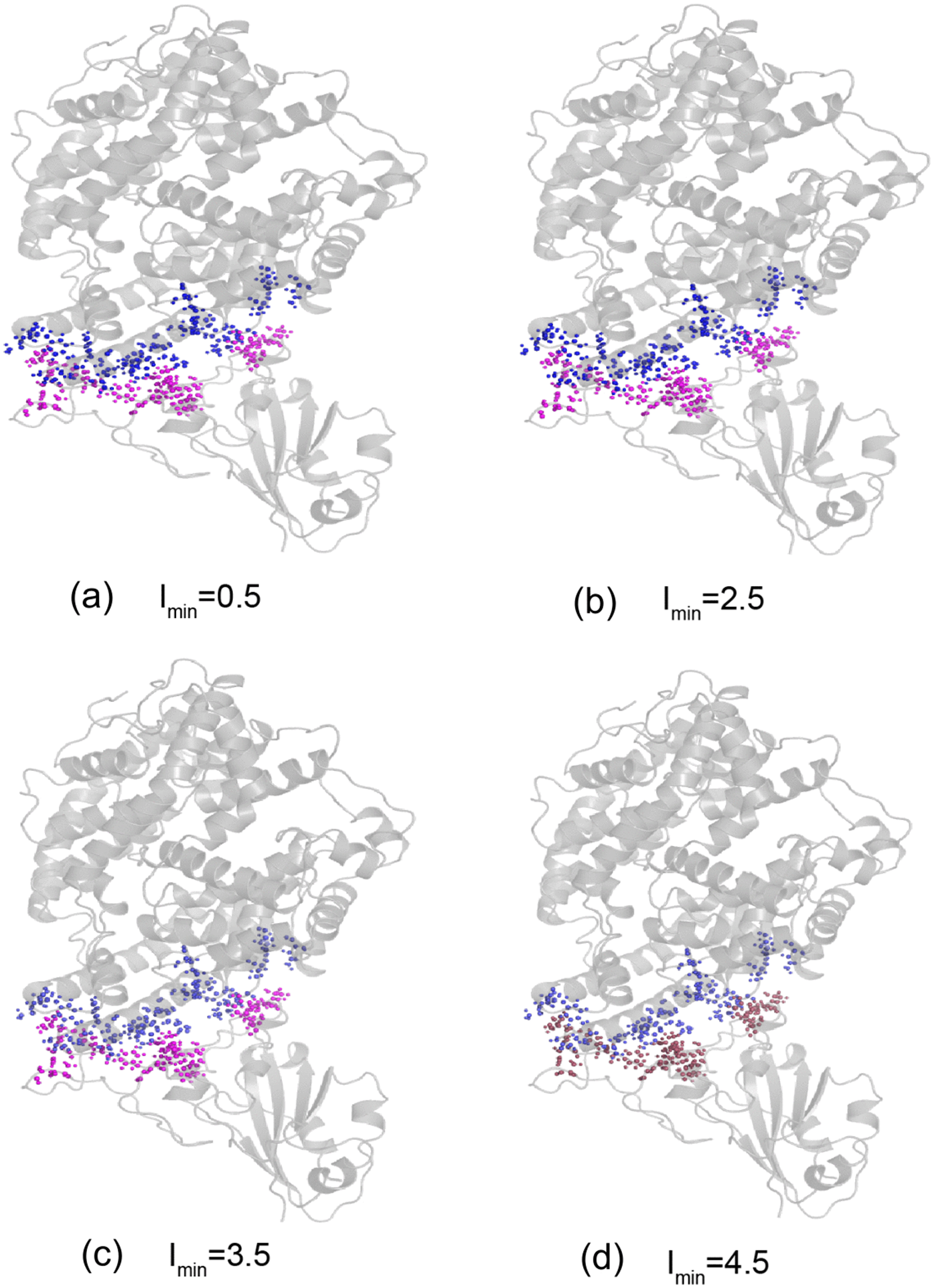}
    \caption{{\bf SARS-CoV2 Interface Clusters:} The PYMOL representation of the interface clusters of SARS-CoV2 complex at various $I_{min}$ at 50$\%$ dynamical stability are presented here. As with Fig \ref{Fig 1} for SARS-CoV1, the nodes corresponding to ACE2 receptor residues are represented in blue, while virus residues are represented in pink. Here too, the breaking up of the largest interface cluster for increasing values of $I_{min}$ is indicative of the percolation-like behaviour. Compared to SARS-CoV1, this SARS-CoV2 complex undergoes a more complex reduction in the size of the interface cluster, which persists up to $I_{min}$$\sim$ 3-3.5. Thus, the interface cluster in this case persists for larger threshold values of edge interaction strength.}
    \label{Fig 2}
\end{figure}

We construct the Protein Sidechain Network (See Methods section) such that the $C_\alpha$ atoms of each residue form the nodes of the network. The non-covalent interactions between any given pair of nodes labeled as $i$ and $j$ are captured by the edge strength parameter $I_{ij}$, expressed as a percentage. The nodes $i$ and $j$ are connected by an edge if their interaction strength $I_{ij}$ is greater than a given threshold value, henceforth denoted as $I_{min}$. Since our analyses first focus on temporally averaged data, we also need a measure for whether an edge exists on average. We therefore define a 'dynamical stability' measure, which gives the probability of finding an edge over several different snapshots. Our network formulation applies not only to individual proteins, but also to protein complexes. {\color{black} We emphasize here that while our starting point comes from MD studies, which have been extensive for SARS-CoV2, we employ the network approach that we have built over decades to take our analyses into a different dimension.} Our techniques allow us to identify network properties, such as the largest connected clusters within the complex and more. As a highlight of our network method, we show the nodes involved in the largest cluster associated with the interface between the viral protein and the human ACE-2 receptor for both SARS-CoV1 and SARS-CoV2 complexes (Figs \ref{Fig 1} and \ref{Fig 2} respectively).

We perform our analysis over this set of constructed networks at various threshold values and lay out our key results in this section. We first present our results on the percolation-like behaviour of both complexes, dynamically averaged for varying values of $I_{min}$, with the predominant presence of interface residues in the largest cluster. We argue that the persistence of such a large interface cluster for higher threshold values in the SARS-CoV2 complex is an indicator of its higher robustness and dominance. These results also indicate the $I_{min}$ values of significance in the network. We then focus on the nature of fluctuations in bond strength during the simulation to further validate the indication of a robust interface. The dynamically averaged cliques and communities obtained at the interface form a connectivity map at the atomic level offering further evidence of a tightly held interface in the case of SARS-CoV2 as compared to SARS-CoV1. We finally discuss the network metrics of a few specific points in the simulation trajectory that highlight the diversity in the conformational landscape of individual snapshots of the complexes in comparison to the dynamical averages.


\subsection*{The Human ACE2 receptor complex forms a more dominant, robust cluster with the SARS-CoV2 spike protein as compared to SARS-CoV1} 
\begin{figure}
    \centering
    \includegraphics[width=\textwidth]{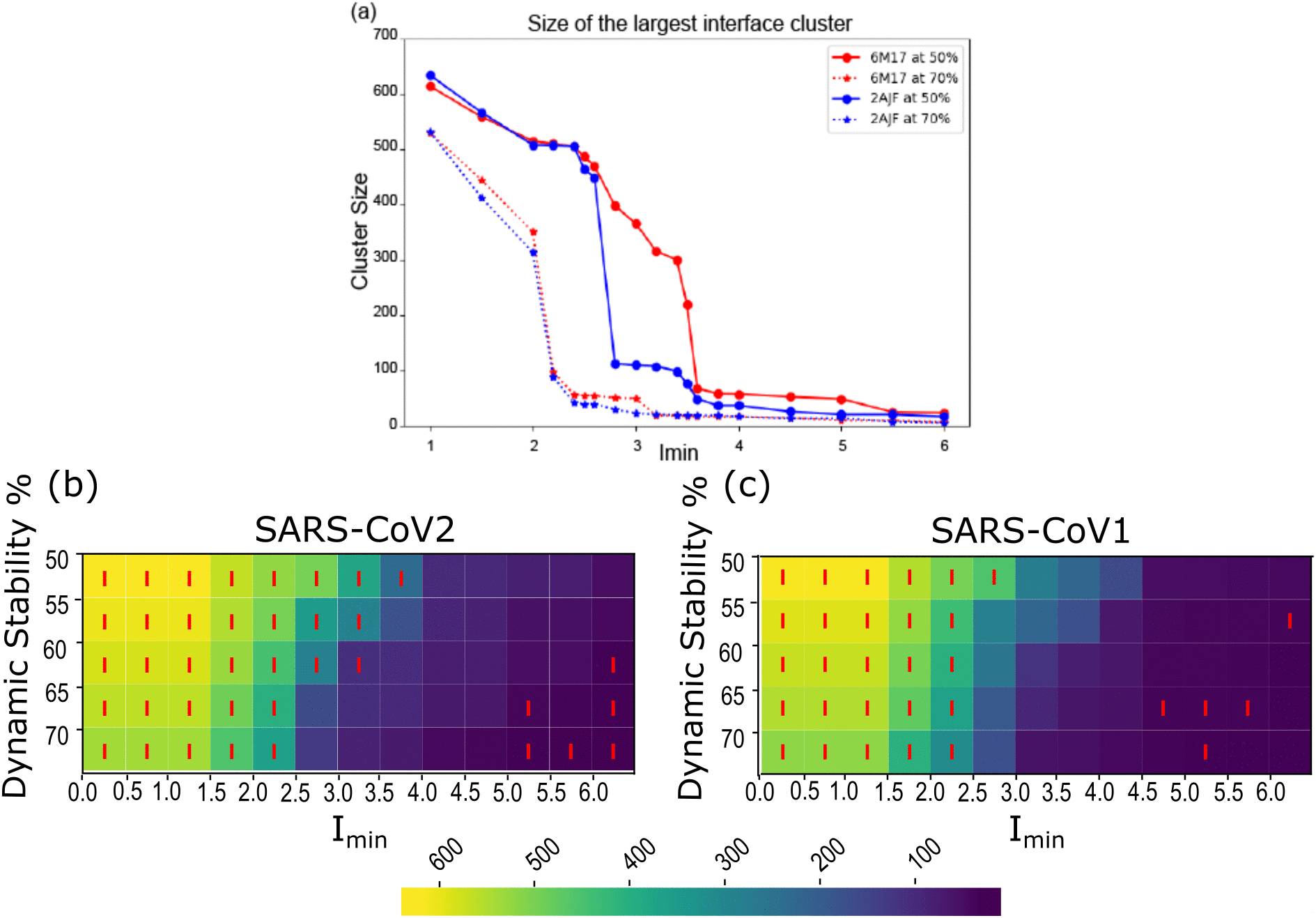}
    \caption{{\bf Mesoscopic view of global connectivity maps within the spike protein – ACE2 receptor complex network :}    Percolation analyses of the largest cluster at the interface reveal a tighter interface in SARS-CoV2 compared to SARS-CoV1. (a) The size of the interface cluster versus the threshold value $I_{min}$ for a bond to exist between nodes of the network show a typical percolation-type curve characterized by a sudden drop. Here we show such curves for two values of dynamics stability (see Methods) of 50$\%$ and 70$\%$, obtained from averaging out MD simulations; a large connected interface cluster persists for larger threshold value in SARS-CoV2.  The variation of the curve for intermediate values of dynamic stability are shown in the heat maps of Figs (b) and (c) for SARS-CoV2 and SARS-CoV1 respectively. The 'I' (orange) indicates regions in parameter space where the largest connected cluster is at the interface; the plots indicate that these interface clusters are more dominant in SARS-CoV2.}
    \label{Fig 3}
\end{figure}

\textcolor{black} {We begin our studies by employing a percolation-based approach on the dynamically averaged spike protein-ACE2 receptor complexes of both SARS-CoV1 and SARS-CoV2. Our results are demonstrated in Fig \ref{Fig 3}}. Crucially, it is seen that in both cases, the largest cluster formed in the network at lower $I_{min}$ values (1 to $\sim$2.75 $\%$) is an interface cluster (IF-cluster, that is, one that contains nodes belonging to both the spike protein as well as the ACE2 receptor), encompassing most of the residues of both the proteins of the complex. However, the IF-cluster persists even at a higher stringency criteria of $I_{min}\sim$3-4$\%$ mainly in the complex of SARS-CoV2 (Fig \ref{Fig 3}(b)) as compared to SARS-CoV1 (Fig \ref{Fig 3}(c)). Further, it is also seen that the IF-cluster continues to persist at higher $I_{min}$  values of dynamic stability in the complex of SARS-CoV2 than in SARS-CoV1. The dynamic stability axis(details provided in the Methods section) gives us an indication of the statistical significance of the network’s connectivity, as its conformations change through time in the MD simulations.  Fig \ref{Fig 3}(a) highlights this aspect of the same data at two values (70$\%$ and 50$\%$) of dynamic stability. At the 50$\%$ stability value, the more intricate nature of the interface percolation transition region in the complex of SARS-CoV2 is sharply contrasted against the near-sigmoidal nature of the SARS-CoV1 curve.  In particular, the size of the largest IF-cluster falls steeply at $I_{min}$ $\sim$2.75 for SARS-CoV1, whereas SARS-CoV2 experiences an additional transition  at $I_{min}$ $\sim$3.5 and the IF-cluster size is noticeably larger at $I_{min}$  up to $\sim$5.5. These observations strongly suggest the robust and dominant nature of interaction at the interface of the two proteins in SARS-CoV2-ACE2 complex. 

\subsection*{Higher order connectivity and its visualization reveal tighter cliques and communities at the interface of the SARS-CoV2 complex as compared to the SARS-CoV1 complex} 
\begin{figure}
    \centering
    \includegraphics[width=\textwidth]{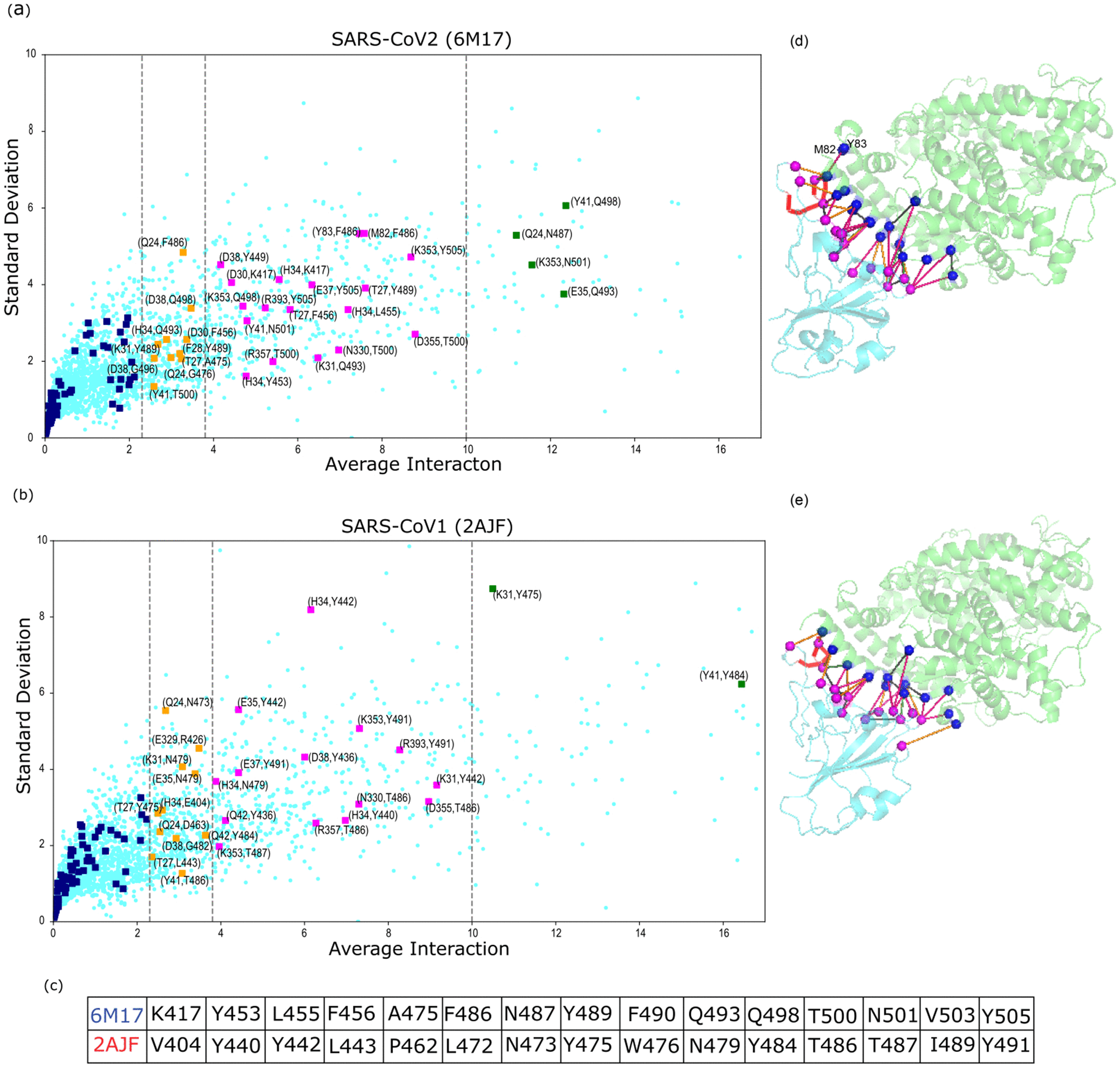}
    \caption{{\bf Interface interactions of the amino acids of the ACE2-Receptor with those of the viral proteins[(a) SARS-CoV2 (PDB ID 6M17) and (b)SARS-CoV1 (PDB ID 2AJF)] complexes:}   This is a zoomed-in view of all edge interaction strengths in the spike-ACE2 complex structure network across the conformational ensemble obtained from the MD simulations. We plot the edge interaction strengths for all pair of residues averaged over all the MD trajectory snapshots against the standard deviation in the interaction strengths. This value signifies the fluctuations from the average edge strength for the corresponding pair. We have also highlighted the interface residue pairs (I,J) where I comes from the ACE2 receptor and J from the spike protein. The vertical grey lines divide the plots to bin the interface edges in four intervals based on average interaction. These intervals are color-coded as navy (0-2.3), orange (2.3-3.8), magenta (3.8-10) and green (over 10). (c) The table below maps the 6M17 spike protein residues to structurally equivalent residues in 2AJF. (d,e) PYMOL representations of protein complexes are also shown alongside, highlighting the interface edges. These plots highlight how the strength of interface edges predominantly stronger in the case of SARS-CoV2. }
    \label{Fig 4}
\end{figure}

Motivated by the global, mesoscopic understanding of the interface from Fig \ref{Fig 3}, we now investigate atomistic details of the network and present our results in Figs 4(a) and 4(b). Here we plot the dynamical average of the interaction strength of each edge against the standard deviation, which indicates the extent of fluctuation. The fluctuations of individual edges vary approximately in the range of 1-6, in both the complexes, except for a few points reaching higher values in the case of SARS-CoV1 complex. Further, most of the fluctuating edges in SARS-CoV2 complex with average interaction strength (2-10) have standard deviation below the value of 4. The IF-edges are colour coded, based on average interaction strength and the interfacial edges with interaction strength greater than 2.3$\%$ are annotated with the interacting residue pairs. Consistent with the previous result, we see that a larger number of interfacial edges in the SARS-CoV2 complex (Fig \ref{Fig 4}(a)) have higher interaction strength (in the 2.3-3.8$\%$ region) as compared with the SARS-CoV1 complex (Fig \ref{Fig 4}(b)). We also present a panel in Fig \ref{Fig 4}(c) that shows sequence alignment of the amino acid residues of the two spike proteins that participate in the interface bonds, thus mapping certain interfacial edges onto each other. The details of the displayed edges and their strengths are compared in S1 Table in the Supplementary Materials.

\begin{figure}
    \centering
    \includegraphics[width=\textwidth]{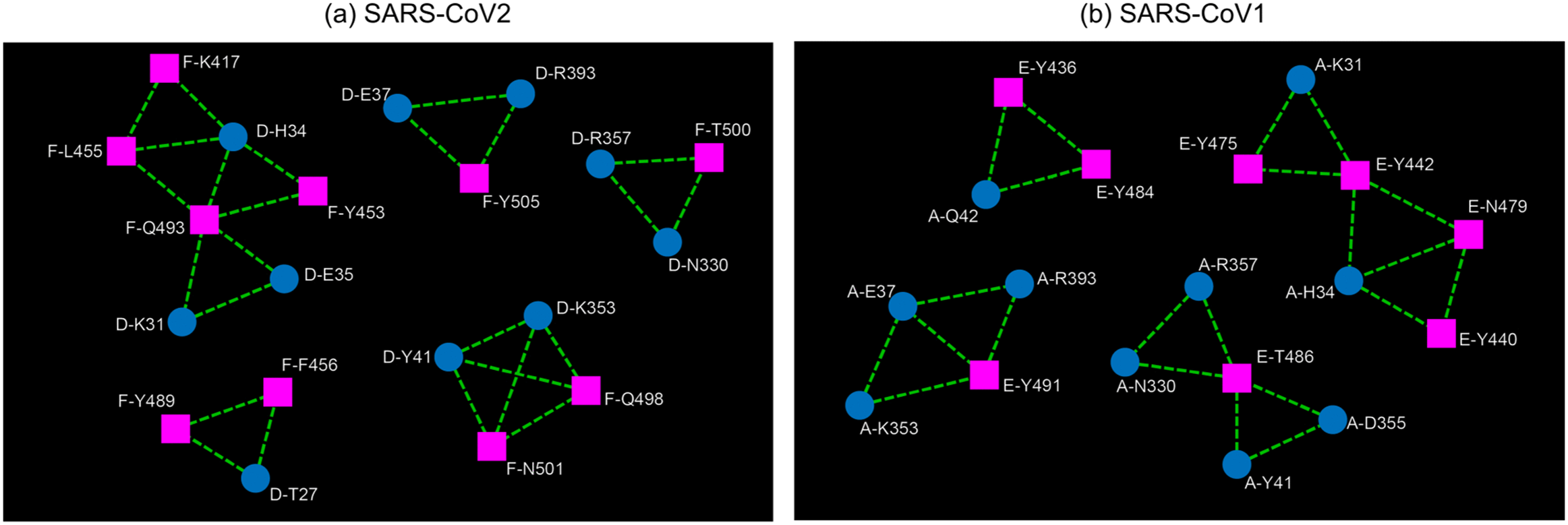}
    \caption{{\bf Interface Cliques for the average network:}  The interface cliques obtained for the averaged network at $I_{min}$ 2.75, 50$\%$ dynamic stability are presented here for the (a) SARS-CoV2 complex and (b) SARS-CoV1 complex. The residues/nodes corresponding to the ACE2 receptor are denoted as blue circles while those corresponding to the virus spike protein are denoted as magenta squares. Node labeling: First letter represents the Chain number in the PDB for each complex; The amino acid residues are shown in one letter code, followed by sequence number. The same representation is used in the figures to follow.}
    \label{Fig 5}
\end{figure}

Interestingly, many of the fluctuating edges at the interface are also part of IF-cliques/communities. Additionally, edges in both complexes with average interaction strength beyond 10 form interface cliques and communities, while also possessing higher fluctuations in edge strengths.  They are likely to play an important functional role by exploring a wider conformational space. Thus, it is useful to compare the two complexes at the next level of higher order connectivity (IF-cliques/communities), and the corresponding spike-receptor binding domain (RBD) residues. The interface cliques of both complexes associated with the average network at $I_{min}=2.75$ are shown in Fig \ref{Fig 5}(a,b). The Figs \ref{Fig 4}(d) and \ref{Fig 4}(e), highlight the locations of the interfacial edges on the 3D representations of the two protein complexes (shown on PYMOL of the reference structure).  The average IF-edges (the color coded strength as in Fig \ref{Fig 4}(a-b)) presented here, indicates the stronger nature of interface interaction between the spike protein and the receptor protein for SARS-CoV2, in terms of the cliques/communities they form. It also underscores the position of functionally important residues like Lys417, Asn501, Leu455, Tyr505, at the community level, many of them are found to be mutated in the variants of interest. 

The interactions of certain aligned spike protein residues of SARS-CoV1 and SARS-CoV2 like Val404/Lys417, Tyr442/Leu455, Pro464/Ala475, Leu472/Phe486, and Tyr484/Gln498 (See Fig \ref{Fig 4}(c)) with the residues of the ACE2 receptor are important to our analysis here. We also draw attention to the edges formed by methionine 82 and tyrosine 83 (ACE2 receptor) with phenylalanine 486 and lysine 417 of SARS-CoV2 with residues aspartic acid 30 and histidine 34 (ACE2 receptor). The loop held by the disulphide bridge in the two spike proteins [SARS-CoV2 (Cys480-Cys488) and SARS-CoV1(Cys467-Cys474)], and their amino acid compositions [CoV2-CNGVEGFNC- versus CoV1-CTPPALNC], has played a crucial role in their interaction with the ACE2 receptor. This loop in SARS-CoV2 is not only longer by one residue, but also contains two flexible glycine residues, in contrast with two highly rigid proline residues in SARS-CoV1. This has enabled SARS-CoV2 to comfortably latch on to the residues methionine 82 and tyrosine 83 of the ACE2 receptor. Furthermore, the interaction between the loop (Cys480-Cys488) of SARS-CoV2 is retained even at higher interaction strength of $I_{min}$=4.5.  Such differences have led to subtle rearrangement of IF-interactions and their average strengths, leading to significant differences in their connectivities.   
 
More explicitly, the accessible amino acid residues of the ACE2 receptor are tightly captured by the RBD region of SARS-CoV2 amino acid residues.  Although the RBD residues of SARS-CoV1 are exposed to the same residues of the ACE2 receptor, it is unable to bind strongly with the residues like Met82, Tyr83 on one face of the complex and to the residues Glu37 and Arg 393 on the opposite face. Further, the residue Lys417 of SARS-CoV2 also plays a significant role by reaching out a long helical stretch (from Asp30 to His34) of the accessible face of the ACE2 receptor, as well as making a strong network of cliques. Such an active connectivity-role is unmatched by the corresponding residue Val404 of SARS-CoV1. Figs 4(a-e) and S1 Table in the Supplementary Materials provide various parameters to support this fact and finally, the explicit atomic details of the interface sidechains are depicted by the PYMOL representation of the two complexes in Figs 4(d,e). 

\subsection*{Higher affinity of SARS-CoV2 to ACE2 receptor as compared to SARS-CoV1 is associated with a higher interaction strength of the IF-cluster} 
\begin{figure}
    \centering
    \includegraphics[width=\textwidth]{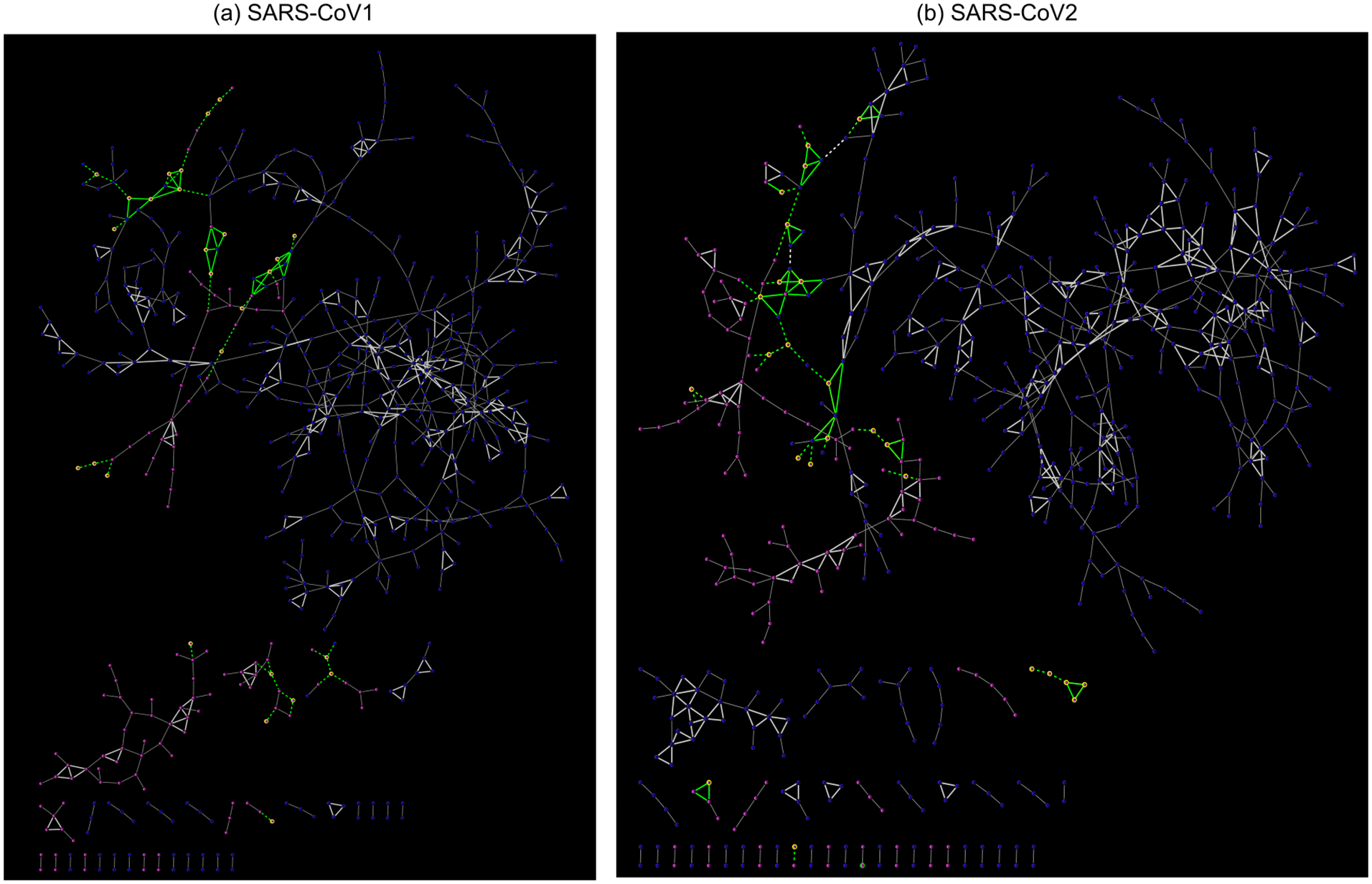}
    \caption{{\bf Largest interface cluster of the average network:} The largest cluster obtained for the averaged network at $I_{min}$ 2.75, 50$\%$ dynamic stability are presented here for the (a) SARS-CoV1 complex and (b) SARS-CoV2 complex. The entire complex is represented by network metrics of edge, clique, community and cluster. The residues connected by green edges represent the interface nodes, and highlight the nature of their extended connection in the SARS-CoV2 complex as compared to the SARS-CoV1 complex. Non-interface cliques are shown in white color.}
    \label{Fig 6}
\end{figure}

As seen in Figs \ref{Fig 6}(a,b), the largest interface cluster at $I_{min}$  $\leq$ 2.75 is dominated by the ACE2 receptor in both the complexes. This situation remains unaltered for higher $I_{min}$ values in the cases of SARS-CoV1 complex, whereas a major part of ACE2 receptor separates from the interface cluster around $I_{min} \sim$ 3.2-3.5 in the case of SARS-CoV2 complex. Further, despite the ACE2 receptor dominating the interface cluster in both complexes at weaker $I_{min}$, we observe clear distinctions between the two cases, in terms of the continuity of the connections at interface. A careful analysis of the corresponding building blocks, cliques/communities in Figs 5(a, b) also exhibit differences. Specifically, these units are well interspersed between the residues of SARS-CoV2 and ACE2 receptor, whereas two out of the three communities in SARS-CoV1 and ACE2 receptor complex are highly biased towards the spike protein residues.  The continuity of the interface connections in the SARS-CoV2-ACE2 receptor complex arises due to the interface communities, which are robust ($I_{min}$  $\geq$ 3.5) and they are stitched beautifully well with strong edges. Furthermore, robustness leads to a strong interface cluster, which separates itself from a major part of the ACE2 receptor. 


This higher affinity of SARS-CoV2 to the ACE 2 receptor is already evident at even the most direct representation in the conventional 3D-pictorial representation (PYMOL) of both complexes for various $I_{min}$ in Figs \ref{Fig 1} and \ref{Fig 2}. This feature continues to be prominent upon our first analysis presented above in the largest cluster transition profiles as function of $I_{min}$  at 50$\%$ dynamic stability as seen in Fig \ref{Fig 3}(a). The stability of the largest interface cluster at higher interaction strengths in SARS-CoV2 complex as compared to SARS-CoV1 complex indicates that information regarding higher binding affinity can be extracted from the size of the largest interface cluster. 

To the best of our knowledge, we have not encountered any studies in literature relating the binding affinity between protein-protein interactions being connected to the interface cluster, with strong non-covalent interactions. This analysis shows that the interface network parameters from the simulation averages indicate a stronger affinity of ACE2 receptor with SARS-CoV2 than with SARS-CoV1. 


\subsection*{Comparing individual conformations shows that while cliques and communities are stable in SARS-CoV2, there exist prominent variations involving key functional residues} We now focus on presenting a few selected conformations of the complex at higher $I_{min}$ values to understand the level of structural diversity in the conformational landscape. Since the SARS-CoV1 complex has already undergone the percolation transition at  $I_{min}$ values greater than 2.75 and the size of the IF cluster is no longer significant, we mainly present the details of interface clusters in snapshots of the SARS-CoV2 complex. We selected the structures (a) Reference, (b) High RMSD, and (c) End, for our analysis, which correspond respectively to the initial equilibrated structure, the snapshot with high RMSD with the Reference, and a snapshot towards the end of the simulations. Comparing individual conformations shows that while cliques and communities are stable, variations involving key functional residues are prominent.  This is evident from the results presented here. The IF-cliques and communities (at $I_{min}$  $\geq$ 3.5) in the three selected snapshots of SARS-CoV2 complex are consolidated in Figs \ref{Fig 7} (a, b, c). It highlights two major points: (a) The amino acid residues in the cliques and communities are mutated (except for Leu 455, until the appearance of currently circulating recombinant variant of interest JN.1), giving rise to variants and secondly, (b) although the cliques are very much similar, they are not identical, clearly showing the diversity of structures at the network level, in the conformational landscape explored during the simulations. That is, a continuous network across the interface in the SARS-CoV2 complex is achieved through robust cliques and communities in the selected snapshots. The network diversity between the three selected structures becomes even more striking with distinctly different conformational states at the IF-cluster level as shown in Figs \ref{Fig 8} (a, b, c). Below we highlight a few points from this perspective.

In the previous section we highlighted the important role of the loop (Cys480-Cys488) in the context of the average behaviors of the SARS-CoV2-ACE2 Receptor complex. This nine-residue loop consisting of the sequence -CNGVEGFNC-, is loaded with structurally influential amino acids. First, it is long enough to latch on to helix 1 of the ACE2 Receptor and interacts with residues like Met82, Tyr83. The two glycine residues (482 and 485) of the loop allow high flexibility, thus accommodating various conformations required at different functional states. For example, one of the cliques of the End structure (Fig \ref{Fig 7}(c)) is made up of N481 and F486 with M82 of the Receptor protein.

A majority of RBD residues of SARS-CoV2 interact with the first 2 helices (ranging from Gln24 to Tyr41) and a few other residues (e.g.,N330, K353, R357, R393) of the ACE2 Receptor. They act as player nodes of the two parties. The collection of cliques and communities from the three selected structures in Figs \ref{Fig 7} (a, b, c) clearly demonstrates the subtle structural diversity brought out by the same set of nodes by regrouping differently. A collection of them and rearrangements with respect to the interface region of the complex is shown in Fig \ref{Fig 9}.

 The integration of the cliques and communities in the interface clusters brings another dimension of diversity as shown in Figs \ref{Fig 8} (a, b, c). The diversity in the interface cluster is brought out by connecting different combinations of cliques/communities. The interface cluster of the Reference structure (Fig \ref{Fig 8}(a)) appears like two arms connected by a hinge region, which is held by the largest community. The IF-cluster regions of the arms are supported mainly by the ACE2-Receptor residues, whereas the hinge region is supported by the RBD residues of SARS-CoV2. By and large, most of the mutations have occurred in IF-cluster. It is noteworthy that until recently significant mutations of the residue L455 were not reported, although it occupies a prominent position in the community and connected to influential residues like K417, Q498. Perhaps it acts as a buffer to maintain the community during dynamical processes. More details of the role of this residue will be presented in the Discussion section. The IF-cluster in high RMSD structure (Fig \ref{Fig 8}(b)) has split up into two clusters. The largest community here is made up of residues such as F498, N501, Y505 of SARS-CoV2, which are connected to mainly the ACE2 Receptor.  The second part of the interface cluster is not dense with communities. However, it has retained the four-residue community with K417 and L455 and a linear connectivity with helix 1 and the residue M82 of the ACE2 Receptor, with the residues of the loop Cys480-Cys488 of SARS-CoV2. Thus, the high-RMSD structure has adopted a strategy of splitting the IF-cluster into two parts, with one of them latching on the Cys480-Cys488 loop and the other, exploring the residues closer to the cleavage site. Finally, the End structure (Fig \ref{Fig 8}(c)) offers a glimpse of a large IF-cluster, connecting the entire length consisting the loop latching to M82, passing through the hinge region (K417, L455, Q493) and reaching the active centre, consisting of important functional residues such as T500, N501, and Y505.

To summarize, the robustness of the interaction between the ACE2 Receptor and the RBD region of the SARS-CoV2 at higher interaction strength is evident from the network metrics of interface cliques/communities, and the interface clusters. More importantly, the conformational diversity depicted by subtle reorganization of the same set of amino acids from the viral protein and the host -ACE2 Receptor is highly impressive. The importance of network parameters in rationalizing some of the mutations and the relevance of conformations diversity at network level, will further be dealt with in the Discussion section.

\section*{Discussion}

 In the Results section we have presented a rigorous analysis of the simulation trajectories of the complexes that the human ACE2 receptor forms with the COVID-19 virus SARS-CoV2 and the earlier coronavirus SARS-CoV1 \cite{DEShaw}.  The connectivity from the one-dimensional edge level interaction leading to three-dimensional higher order topological units, culminating in globally connected clusters are systematically tracked through detailed connectivity maps. Furthermore, the dependence of global connectivity on edge weights has unraveled intricate changes that undergo in the dynamical state.  The higher dynamical stability of SARS-CoV2 complex than SARS-CoV1 complex is  observed from  the percolation behavior of simulation averages, as well as from the specific edge strengths, their fluctuations, and from the topologically higher order connectivity metrics. Apart from the average behavior, the analysis of individual snapshots of the SARS-CoV2 complex has provided a landscape of conformations, with subtle differences in their cliques, communities, and the IF-clusters, which are not captured by conventional parameters such as root mean square deviation (RMSD) of the overall backbone structure. Needless to mention that such a connectivity map can be useful in many ways, such as locating the position and the topological status of relevant residues of the viral protein as well as the interacting residues of the ACE2 receptor in the global context. For example, such a map can assist in placing the mutations of variants on the global map or in designing drugs and vaccines.

Here, we underscore the importance of the connectivity map and the network topological metrics, which would have been difficult to rationalize through direct interface connections alone. Firstly, we have located a majority of the mutations in the receptor binding domain of SARS-CoV2, at the interface of the complex with ACE2 receptor, from the sidechain non-covalent connectivity maps. Based on the results presented in the previous section, below we present our analyses on its implications to understanding Sequence-Strucrure-Functional correlation.

\subsection*{The 3D-map generated from the PScN complex allows pinpointing of mutations}  Since the emergence of the COVID-19 pandemic at the end of 2019, extensive sequencing of SARS-CoV2  containing 1273 amino acids has been carried out.  Some of the new articles (See \cite{Guruprasad2023-ef,N2,N3,Cao2023,Wang31122024,Liu2025} and the references there-in) provide a wealth of information on mutants. \ref{S1 Fig} Fig and \ref{S2 Fig} Fig in which the details of various domains like S1, S2 and their cleavage and several other important regions/domains are presented in the context of viral entry into the cell. Understandably, the focus is mainly on the S1 domain (particularly on RBD) and the cleavage site of S1-S2 domains, since various dynamical functions, such as the affinity with ACE2-Receptor, antigenicity , the detachment of S4 domain, are performed by sub-domains of S1. The crucial first step of attack is the binding of RBD with the ACE2 receptor. New active variants are still emerging, and one needs to understand the complete mechanism of various aspects such as intensity of infection, escape mutations, rate of transmissibility and so on. A significant number of structures of the complex with ACE2 receptor have also become available from X-ray and cryo-EM studies. As expected, most mutations are at the binding site. Although the direct interactions between the residues of SARS-CoV2 and ACE2 receptor are obvious from the structures, there are several of them which can be detected only through the network representation. Here we highlight some of the mutated residues reported in literature, on our connectivity map of the IF-Cluster, generated at the edge interaction strength $I_{ij}$ $\geq$ 3.5 of the Reference structure, as shown in Fig \ref{Fig 8}(a). 

An extensive diversification of Omicron sublineages has taken place in the past two years. Among them, the subvariants BQ.1 and XBB.1 were reported to be the most resistant SARS-CoV2 variants. Recombination of several sublineages spreading simultaneously, have led to complex-recombinant variants, making it difficult to track the lineage of off-springs through conventional approaches. Extensive experimental studies and analyses are being conducted to address the problem of induced convergent evolution of Omicron-RBD domain\cite{Cao2023, Bloom2023}. \ref{S3 Fig} Fig (borrowed and presented as is from  Reference \cite{Cao2023}) provides a lucid picture of evolutionarily conserved mutations in the emergent variants. In addition, other variants of SARS-CoV2 BA.2.86 lineage  \cite{Kaku2023} such as JN.1(L455S mutation), and Flip Variant with mutations (L455F; F456L) are currently posing challenges with high transmissibility/binding affinity, in several parts of the globe. Further, the fast genomic mutations in SARS-CoV2 from immune compromised patients are also providing information for further exploration \cite{FUTATSUSAKO2024109597}. With this background of constantly evolving variants of concern/interest, we provide a perspective from the structure network of SARS-CoV2 complex with ACE2 receptor.

\begin{figure}
    \centering
    \includegraphics[width=\textwidth]{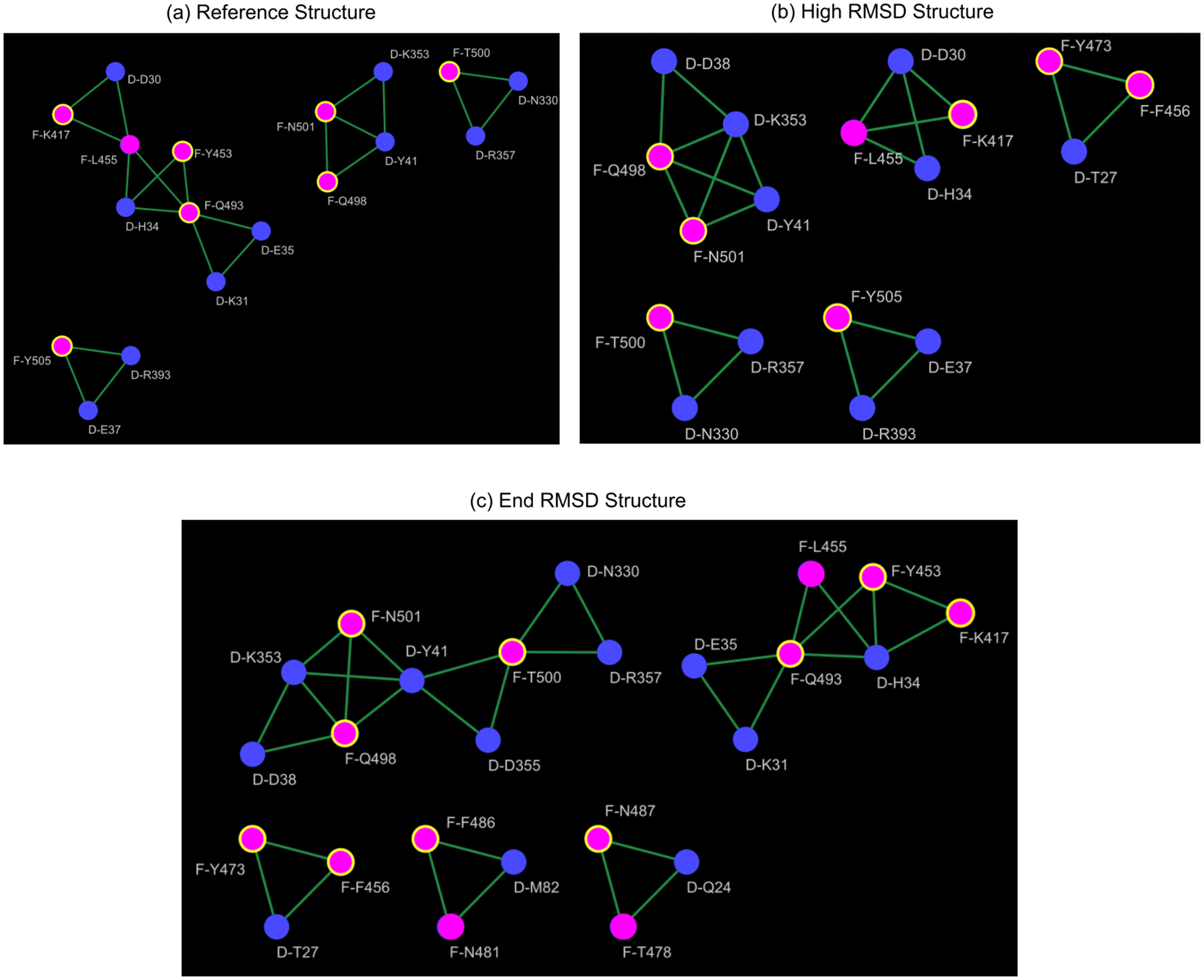}
    \caption{{\bf Snapshot Interface Cliques:} The interface cliques/communities of the SARS-CoV2 complex in three selected snapshots ($I_{min}$ =3.5) are displayed here for the (a) Reference structure, (b) High RMSD and (c) End RMSD respectively. The mutated residues in variants  are represented with a yellow outline on the nodes. The cliques obtained here as well as the nodes involved are similar to those obtained for the average network in Fig \ref{Fig 5} but are significantly different highlighting the diversity of conformations explored during the simulation time.}
    \label{Fig 7}
\end{figure}

\begin{figure}
    \centering
    \includegraphics[width=\textwidth]{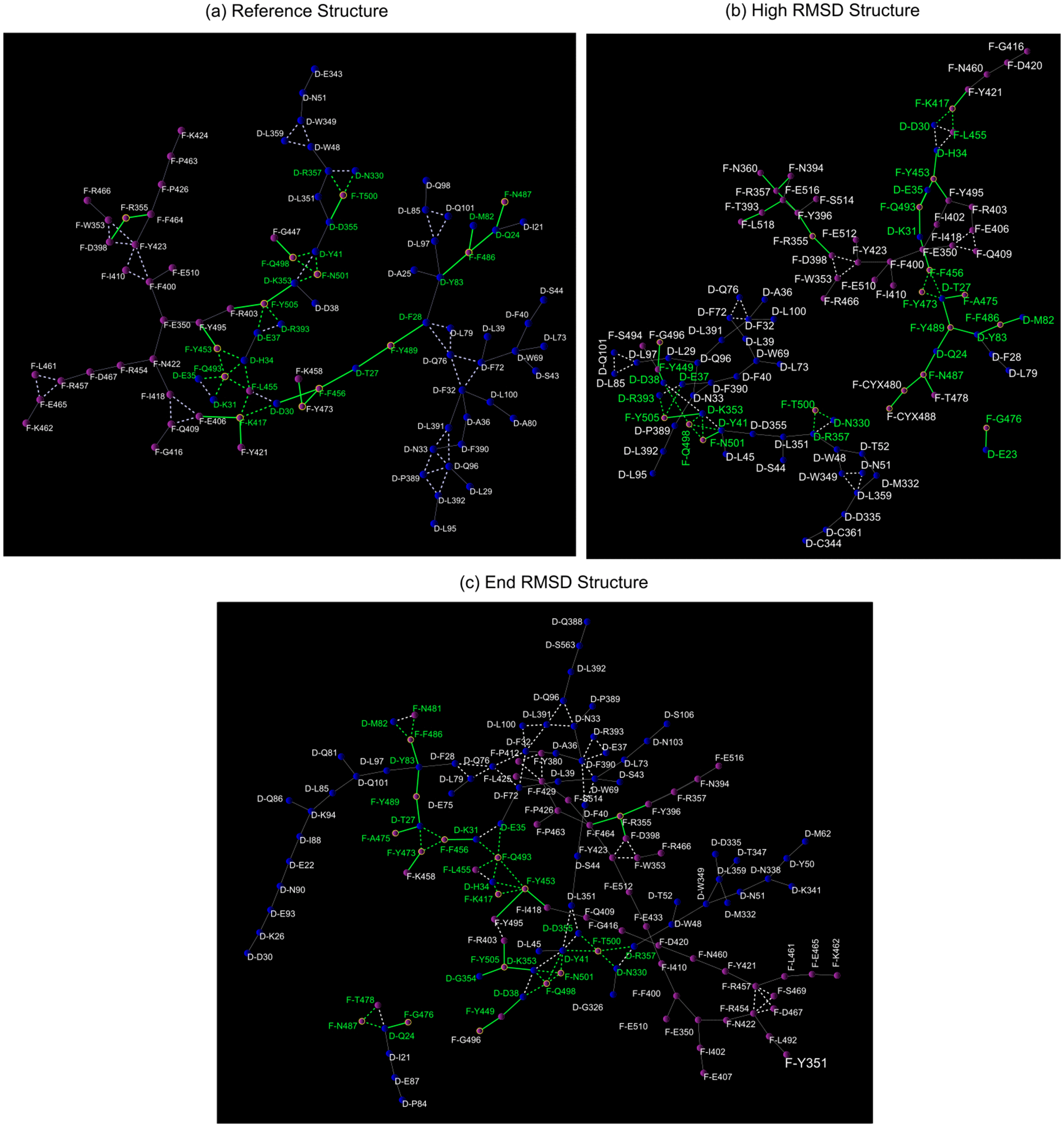}
    \caption{{\bf Snapshots of Interface Clusters:} The IF-Clusters of SARS-CoV2:ACE2 receptor complex of three selected snapshots ($I_{min}$=3.5) are displayed here for Structures (a) Reference, (b) High RMSD, and (c) End RMSD. The ACE2 receptor nodes (residues) are represented in blue spheres. The SARS-CoV2 nodes are shown in pink spheres and are encircled by orange colour, if they are reported to be mutated (from Wuhan to Omicron Variants). Dotted green lines indicate interface clique connections and solid green lines indicate interface (IF) connections that do not belong to any clique. Dotted white lines indicate non-IF cliques and thin lines indicate non-IF connections that do not belong to a clique.   }
    \label{Fig 8}
\end{figure}

\begin{figure}
    \centering
    \includegraphics[width=\textwidth]{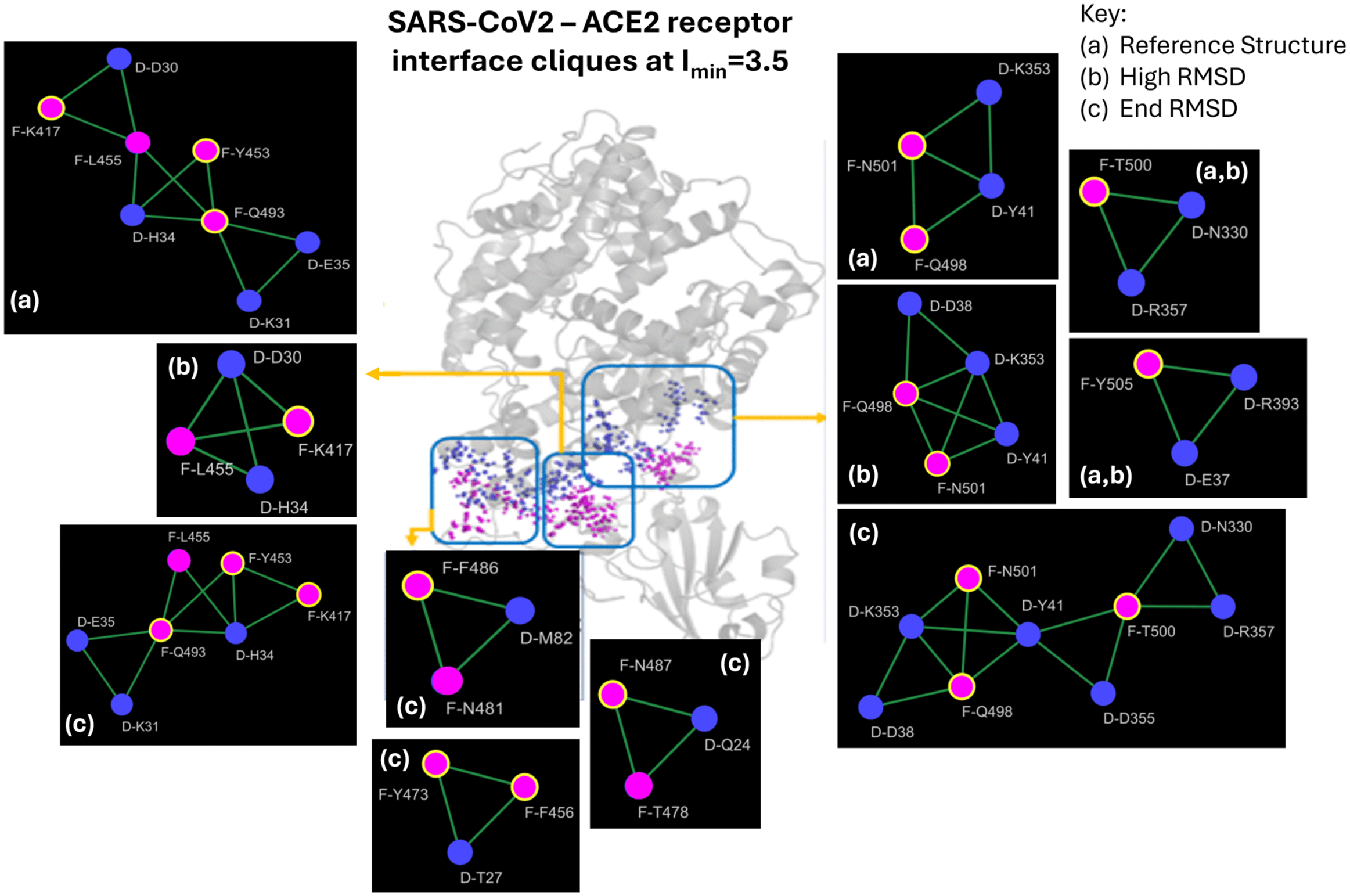}
    \caption{{\bf Relative positions of Interface Cliques:}  The interface cliques in the three selected snapshots ($I_{min}$ =3.5) are represented over the PYMOL structure corresponding to the SARS-CoV2 complex, denoting their positions along the interface. }
    \label{Fig 9}
\end{figure}

 Figs \ref{Fig 7}, \ref{Fig 8}, and \ref{Fig 9} clearly indicate that most of the mutations are embedded in cliques/communities, which are generally parts of the largest IF-Clusters. Here we compare a few of these residues with the set of mutated residues presented on the Y-axis of \ref{S3 Fig} Fig, which are  implicated to be involved in convergent evolution of SARS-CoV2-RBD region. For example, R493 of BA.5 variant has reverted back to the original Wuhan strain Q493 and interestingly, in two simulation snapshots (Figs \ref{Fig 7}(a), \ref{Fig 7}(b)), it occupies a prominent position in the community. Similarly, Q498 also occupies a prominent position in the community (including the key residue N501) in all the three snapshots (Figs \ref{Fig 7}(a-c)). Although Q498 is not listed in \ref{S3 Fig} Fig, it is attached to G447 in the IF-cluster (Fig \ref{Fig 8}(a)) and its sequential neighbours G446 and V445 that are listed in \ref{S3 Fig} Fig, have possibly undergone convergent evolution to serine and proline residues respectively, which can greatly influence the conformational landscape.
 
Finally, the JN.1 and the Flip Variant with mutations (L455F; F456L) are interesting cases to view from structure network perspective. Until recently, significant mutations in L455 were not reported. Looking at it’s crucial position in the community containing K417 and several other key residues, we were intrigued by its resistance to mutation. Now, in JN.1 it is mutated to a small serine residue. The flip variant is even more interesting, by swapping the leucine residue with the obligatory sequence neighbour F456, essentially maintaining the leucine environment in the vicinity. Experiments have indicated that the SARS-CoV2  Omicron JN.1 mutants such as KP.2 and KP.3, with recurrent spike mutations of residues like R346T, F456L, and T572I, Q493E exhibit growth advantages \cite{Wang31122024}. This information can also aid in the design of new Covid-19 vaccine boosters. Thus, the connectivity map can provide a wealth of information in understanding  complicated mechanisms such as allostery, paths of communication, evolution of sequence-function relationships \cite{rf10} or inferring the effect of mutations from a global structural perspective.


In the second part of this section, below we focus on understanding the rationale behind the efficacy of designed ACE2 decoy receptors to effectively neutralize the constantly evolving coronaviruses, by mapping the experimental results on the SARS-CoV2-ACE2 receptor networks that we have generated in this study. 

\subsection*{The global connectivity map can potentially assist in designing ACE2 decoy receptors to neutralize the Coronaviruses}  

 As a treatment towards SARS-CoV2 infection, monoclonal antibodies(mAbs) and their cocktail forms are designed to combat viral evasion. However, the evolution of Omicron and other variants of SARS-CoV2, such as BQ.1, XBB, JN1 and its variants have managed to evade neutralization due to their escape mutations. Thus, the concept of engineered ACE2 decoy receptors has been explored to neutralize SARS-CoV2 virus evasion and the clinical studies have gained momentum in recent times. 

\begin{figure}
    \centering
    \includegraphics[width=\textwidth]{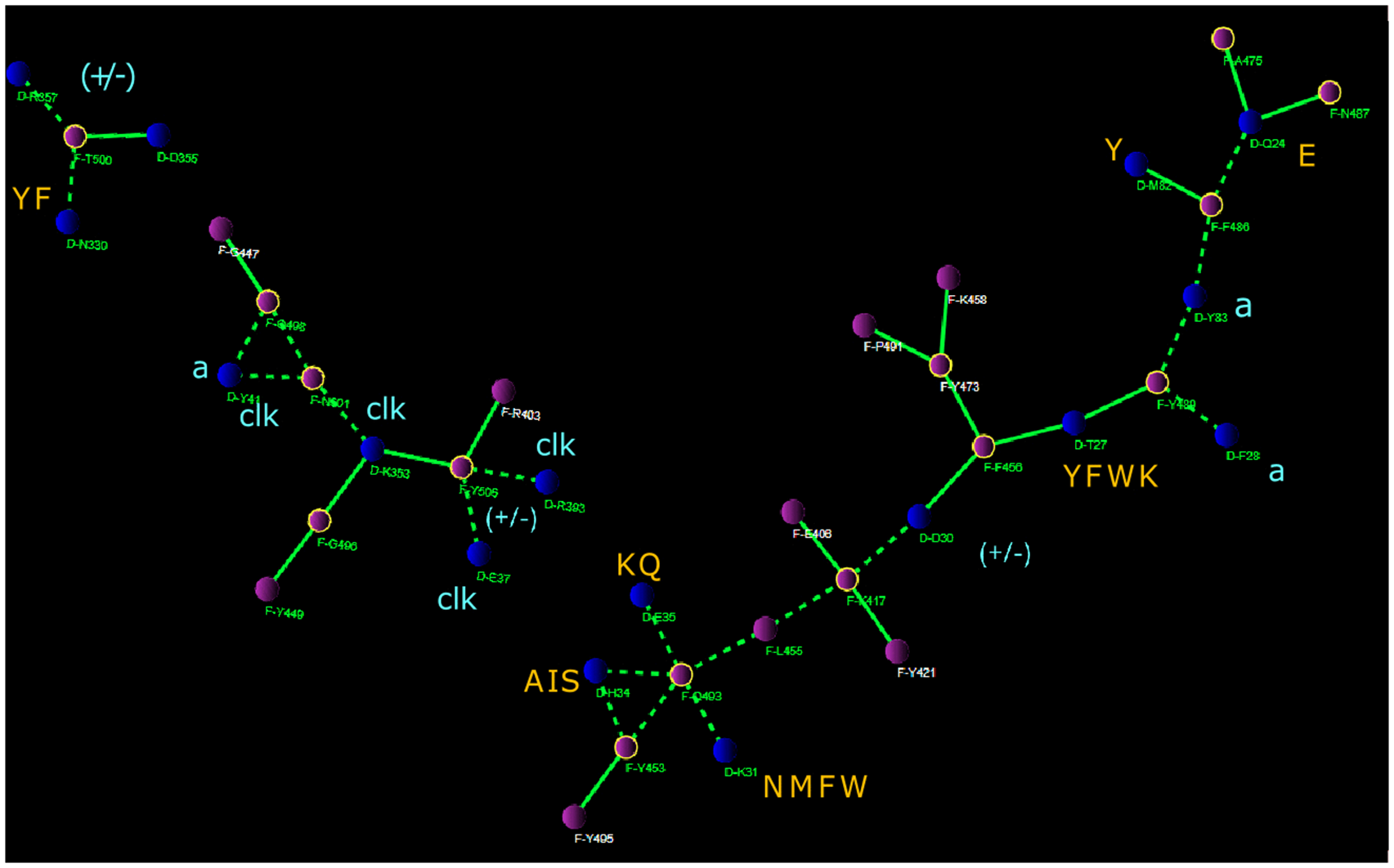}
    \caption{{\bf Designed mutations in the ACE2 receptor interacting with SARS-CoV2:}  Details of individual nodes, which increased the binding affinity between SARS-CoV2 and the ACE2 receptor [Reference \cite{Arimori2022-ls}, adapted in \ref{S4 Fig} Fig] in the engineered decoy ACE2-receptor, are depicted on the IF-Cluster region of the SARS-COV2 Reference structure (Fig (8a)). They are shown in capital single letter amino acid code in yellow/orange color. Some of the residues in the IF-Cluster, which did not show any affinity enhancement upon mutation, due to their chemical properties, are marked as: (a) aromatic residues, (+/-) the charge complementarity residues, and (clk) the residues-participating in cliques.  }
    \label{Fig 10}
\end{figure}

    Various experimental and computational strategies \cite{Chan2023,Huang2020,Chen2021} are used for the design procedure. One of the concepts is to identify the amino acid residues of the human ACE2 receptor, involved in binding (directly or indirectly) to receptor binding domain (RBD) of SARS-CoV2 spike protein and design the ACE2 receptor residues for better binding. In this context, we believe that our map can assist in predicting the viable residues, specifically those which make indirect contact that are difficult to identify by conventional methods. To be precise, we can investigate the ACE2 receptor residues which contribute to better binding through cliques, communities, and part of strong interface cluster (IF-Cluster). As an example, we examine the ACE2 receptor residues which have shown enhanced affinity upon mutations \cite{Arimori2022-ls}, which are presented on the ACE2 receptor-SARS-CoV2 interface complex in \ref{S4 Fig} Fig. Here we have indicated several of the mutations of ACE2 receptor on the IF-cluster of the complex on the Reference structure snapshot (Fig \ref{Fig 8}(a)).   We notice a few interesting correlations of the affinity enhancement due to mutations in the ACE2 receptor, on our highly connected IF-Cluster. Specifically, the important residues of the first 2 helices of the ACE2 receptor prefer aromatic/planar/ sidechains[T27(W,Y,F,K); K31(N,M,F,W); E35(K,Q)] as seen in Fig \ref{Fig 10}. The residue T27 of the ACE2 receptor in this interface cluster is sandwiched between two aromatic residues (F456 and F489) of SARS-CoV2 and the stability around this region is enhanced by the replacement of T27 to aromatic residues. In fact, F456 and F489 of SARS-CoV2 are additionally connected to aromatic residues Y473 (of SARS-CoV2) and F28 (of ACE2 receptor) respectively. In addition, F28 is part of a community of two-3 residue cliques of ACE2 receptor and L79, the direct connection of the first clique shows enhanced binding when mutated to F and T   and two residues in the second clique [F32, F72, Q76] are aromatic residues in the native sequence.  Furthermore, the mutation M82 to Y enhances its interaction with F486, a residue part of the crucial loop (C480-C488) in SARS-CoV2. Similarly, the middle region of the interface network, containing K417 of SARS-CoV2 stabilizes by mutations of K31 of ACE2 receptor to aromatic residues (F,W). On the other hand, the aromatic residues (represented as (a) in Fig \ref{Fig 10}) such as F28, Y83, and Y41 in the IF-Cluster of the native ACE2 receptor have resisted any mutation for increased affinity towards the SARS-CoV2. Of course, some of the residues involved in charge neutrality and are part of cliques, represented as (+/-) and (clk) in Fig \ref{Fig 10} are also not mutated.  
    
Thus, in principle, the engineered ACE2 receptor enhances the binding affinity of the complex by connecting the entire stretch of the interface mainly through the interactions of the aromatic residues. Reinforcement of this hypothesis, at the clique level is evident from the three selected snapshots (Fig \ref{Fig 9}).  IF-cluster maps for the other two snapshots (High RMSD and End structures in Figs \ref{Fig 8}(b) and \ref{Fig 8}(c)), are similar to the IF-Cluster network of the Reference structure (Fig \ref{Fig 8}(a)), as shown in \ref{S5 Fig} Fig and \ref{S6 Fig} Fig. However, structural diversity in IF-cluster is achieved by subtle differences in edges/cliques/communities, which are reorganized with almost the same set of nodes from SARS-CoV2 and ACE2 receptor complex. The END structure shows an interesting behavior, with the entire interface stretch connected. In this structure, all the three functionally crucial nodes of SARS-CoV2 (T500, N501, and Y505) are in the same community , interacting with the ACE2 receptor [residues from(second helix(D38, Y41), the loop region (K353 to R357), and N330], whereas the  T500 of SARS-CoV2 interacting with N330 and R357 splits apart from the main IF-cluster in the Reference and the High-RMSD structures. Thus, in this section, we postulate that mutations in SARS-CoV2 or in the ACE2-receptor results in the redistribution of different structural landscapes at the side chain interaction level, with the redistributions being dependent on the cumulative effect  of  mutations in a given variant of SARS-CoV2 and with the ACE2 receptor (designed or the native ones from different species).

\section*{Conclusion}
In summary, the network representation of all non-covalent interactions in the protein
structures of the complexes that the human ACE2 receptor forms with the COVID-19 virus SARS-
CoV2 and the earlier coronavirus SARS-CoV1 performed here, provides a three-dimensional map,
which can be quantified in terms of network metrics such as cliques, communities and clusters. In
other words, the connectivity from the one-dimensional edge level interaction leading to three-
dimensional higher order topological units, culminating in globally connected clusters, are
systematically tracked through detailed connectivity maps. Here we have specifically captured the
interface interactions, their strengths and fluctuations at the edge level. These features of simulation averages and the percolation behavior, along with higher order topological metrics  have contributed to our understanding of global connectivity at the atomic level. These network based analyses have clearly
exhibited the dynamically stable behaviour of SARS-CoV2 complex than the SARS-CoV1 complex,
which could not have been captured by conventional structure analysis parameters like RMSD.

Based on our analyses of individual snapshots, we find that while the amino acid residues
participating at the interface edges, cliques, and communities of SARS-CoV2 and the ACE2 receptors,
are by and large common as seen from the simulation averages, the diversity is brought out at the
interface cluster level. Thus, we underscore the subtle reorganization of the connectivity of different
units leading to diversity in the conformational landscape, in which the conformations can
dynamically switch from one to another, facilitating specific functions.  {\color{black} In this regard, we speculate on connections with free energy landscapes associated with proteins and conformations, and note that this description takes into account not only enthalpy associated with the connectivity bonds underlying our percolation approach, but also entropic considerations; a closer analysis of these connections could offer insights on the crossover from behavior at the microscopic bond level to that of the macroscopic protein network.}

We have demonstrated the utility of the connectivity map in pinpointing the position and the
topological status of residues involved in mutation/evolution of the viral protein as well as the ACE2
receptor residues, which aid in designing decoy ACE2 receptors for combating viral attack. {\color{black} In fact, our approach could perhaps directly connect with allosteric therapeutical strategies that suggest targeting the ACE2-receptor interaction as a promising route for antiviral drugs\cite{Dutta2022}.}

In conclusion, the network approach presented here to characterize COVID-19- Human ACE2 receptor
can be adopted to understand the mechanism of host-pathogen interaction in general, since the
connectivity map can provide a wealth of information in understanding complicated mechanisms
such as allostery, paths of communication, evolution of sequence-function relationships and so on. A
better understanding of fundamental principles is required to gain insights to the folding process of
proteins in the cell environment \cite{Rose}. In parallel, the AI based AlphaFold \cite{Jumper2021, Abramson2024} can model protein structures from sequences with high accuracy and unimaginable speed. With all
the current advances, an exciting future can be expected in the understanding of the role and
control of proteins in the functioning of cells.


\section*{Methods}
\subsection*{Dataset:} The data for 10 $\mu$s molecular dynamics simulation trajectories of the human ACE2 receptor in complex with the receptor binding domains of the spike protein of the SARS-CoV1 coronavirus (PDB: 2AJF) as well as the SARS-CoV2 coronavirus (PDB: 6M17) were obtained from DE Shaw Research \cite{DEShaw}.  {\color{black} We remark here that the MD simulation data from DE Shaw does a fine job in taking into account various effects such as those of glycosylation, and commandingly handles long-time runs, thus offering an excellent starting point for our network analyses. }The SARS-CoV1 complex has 777 amino acid residues while the SARS-CoV2 complex has 775 residues. This data was further processed to extract 99 snapshots from the trajectories sampled at equal intervals of time for each complex. These two sets of 99 snapshots each form the basis of our analysis, and the input to PSN-Ensemble - the software which extracts network features from the simulation ensemble \cite{Bhat}. All our data is primarily visualized using PYMOL and Cytoscape, with some figures being generated on Adobe Illustrator and InkScape. 

\subsection*{Construction of Protein Structure Networks and their dynamics:} We have performed the sidechain non-covalent connectivity network analyses on the structures of the complexes of SARS-CoV1 and SARS-CoV2(COVID-19) spike proteins with Human ACE2 receptor. The network analyses are performed  on (i) the average connectivity matrix from the simulation snapshots and (ii) on selected snapshots which show commonality and differences in their network parameters such as cliques, communities and large clusters, with focus on the interface region between the ACE2 receptor and the RBD of the spike proteins. Focusing first on the average connectivity matrix, it is obtained based on two inputs - the threshold edge interaction in each network snapshot and the threshold statistical significance of each edge interaction over all the snapshots. A brief summary of the evaluation of the average network based on these parameters is provided below. The detailed methodology is provided in earlier papers/reviews \cite{Frontiers,jcim,Bhattacharyya2016-qt,Bhat,KANNAN1999441}.

Each snapshot is translated to a network where the non-covalent connectivity within a protein structure as well as across the complex are considered as edges, and the nodes can be either the backbone atoms such as C-$\alpha$ atom or an amino acid residue(including all atoms of the sidechain) of the polypeptide chain. Here we have considered the protein sidechain network (PScN), to capture the subtleties of sidechain conformations through weighted edges as given in Equation \ref{eq1} \cite{KANNAN1999441, Sathyapriya}. The interaction between two nodes $i$ and $j$ is given by 
\begin{align}
    I_{ij}=\frac{n_{ij}}{\sqrt{N_iN_j}}\times 100\label{eq1}
\end{align}
where $n_{ij}$ is the number of atom pairs between two given residues $i$ and $j$, within a distance cutoff of 4.5 angstroms, and $N_{i}/N_{j}$  are normalization values defined as the maximum possible number of contacts that the residues $i$ and $j$ can make across a nonredundant database. A weighted PScN is constructed by comparing $I_{ij}$ to the user-defined threshold $I_{min}$ to obtain a binary matrix of connectivity. 

\begin{figure}
    \centering
    \includegraphics[width=\textwidth]{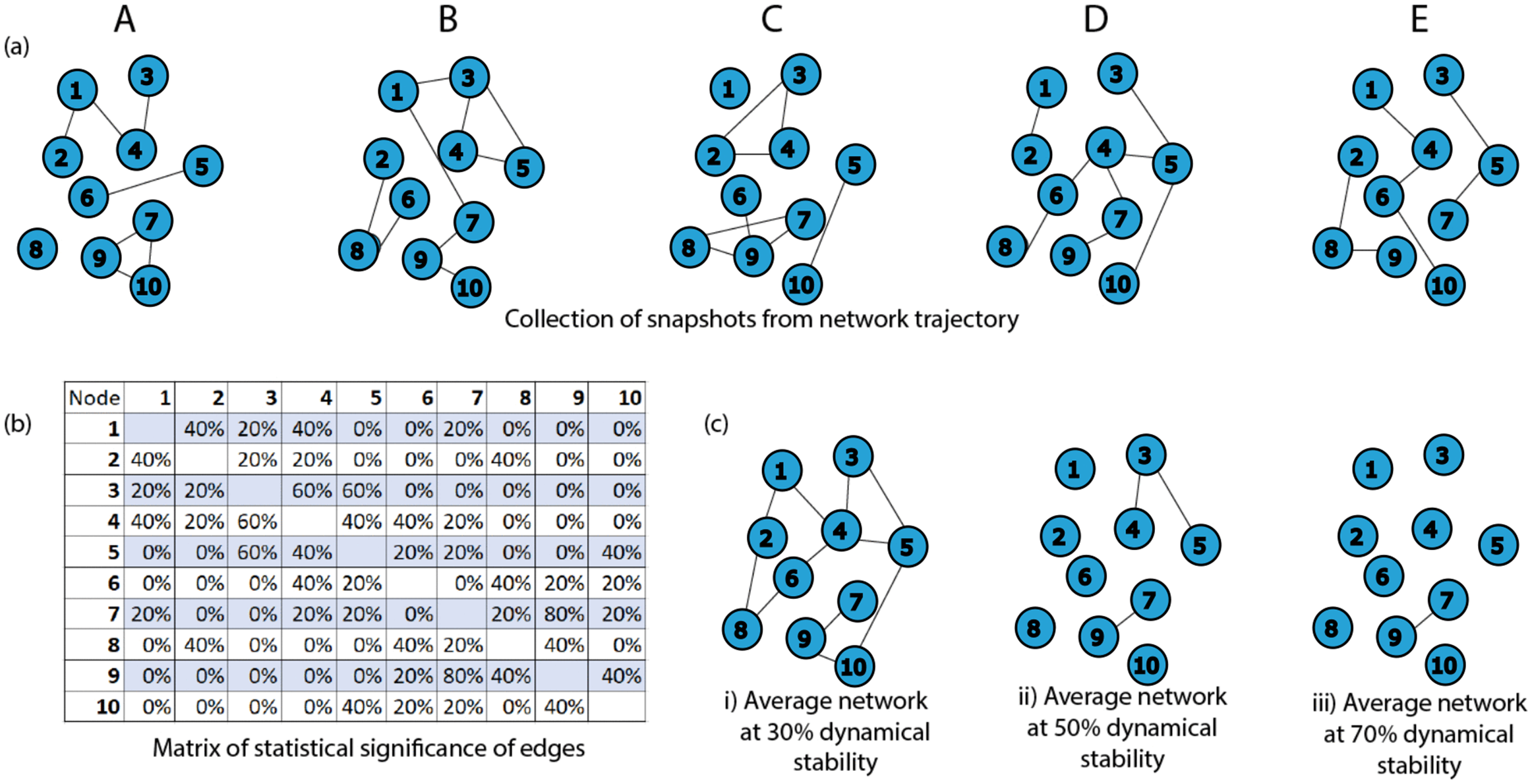}
    \caption{{\bf Understanding Dynamic Stability:}  The above figure offers a brief explanation for the concept of dynamic stability, which is a measure of statistical significance of a certain edge interaction over the entire period of simulation. We elucidate this idea using an example here. Consider a network whose edge interactions are being modified during the course of a certain amount of simulation time. Let the networks A-E in (a) be a collection of snapshots from this simulation trajectory. In (b), we represent the statistical significance of the edge between any two nodes in the network in the snapshots A-E as a matrix. That is, each entry in this matrix is the percentage of snapshots in which a particular edge is present. Here, dynamic stability is threshold value denoting the minimum statistical significance of each edge in the network. Therefore, the average network can be obtained given a certain value of dynamic stability. Such average networks for different values of dynamic stability are shown in (c). }
    \label{Fig 11}
\end{figure}

The average network as defined over all such snapshots requires the second user-defined parameter that we call dynamic stability. Consider the elementary example in Fig \ref{Fig 11}. Fig \ref{Fig 11}(a) shows a collection of networks obtained from snapshots across a simulation trajectory where each edge has a weight of 1. Fig \ref{Fig 11}(b) is a matrix showing the percentage of the total snapshots in which each edge appears. Thus, it is a matrix of statistical significance of the particular edge across the snapshots. Dynamical stability thus specifies the threshold for this value of statistical significance for each edge in the average network. The average networks for three such dynamical stability values are shown in Fig \ref{Fig 11}(c).

\subsection*{Percolation Transition Profile:} Based on the atomic connectivity metrics $I_{ij}$ given in Equation \ref{eq1}, the threshold  $I_{min}$ is expressed as a percentage.  For protein structures, it is shown that the profile of  $I_{min}$  versus the size of the largest cluster thus obtained exhibits a sigmoidal curve, analogous to the percolation phenomenon \cite{brinda,deb}. That is, for a fixed value of dynamic stability and low values of $I_{min}$, the network is dense, leading to a maximally connected cluster that spans system size by upto 80-90$\%$. For higher values of the threshold, the largest cluster size consists only of the very strongly connected edges, which do not span system size. The resulting value of $I_{min}$ with a steep transition in size of the largest cluster is called the percolation point. This point aids us in selecting a normalized connectivity value, which is an optimal nucleation point. The addition of edges with connectivity lower than the transition point will enhance the size of the largest cluster, while setting $I_{min}$ higher than the value at the transition point, yields stronger localized clusters \cite{Frontiers}. Although such a profile is seen largely in globular protein structures, factors such as the interaction of protein-protein complexes can influence the details of the transition profile. In the current study we have observed such changes and have found it to be relevant to distinguish SARS-CoV1 and SARS-CoV2 binding with ACE2 receptor, which have been elucidated in the Results section. 

The clusters, cliques and communities can also be evaluated for individual snapshots without the application of any averaging technique. We perform these analyses for selected snapshots to highlight the effects of connectivity in the complexes.

%
%
%

\setcounter{figure}{0}
\renewcommand{\figurename}{}
\renewcommand{\thefigure}{S\arabic{figure}}

\begin{figure}
    \centering
    \includegraphics[width=\textwidth]{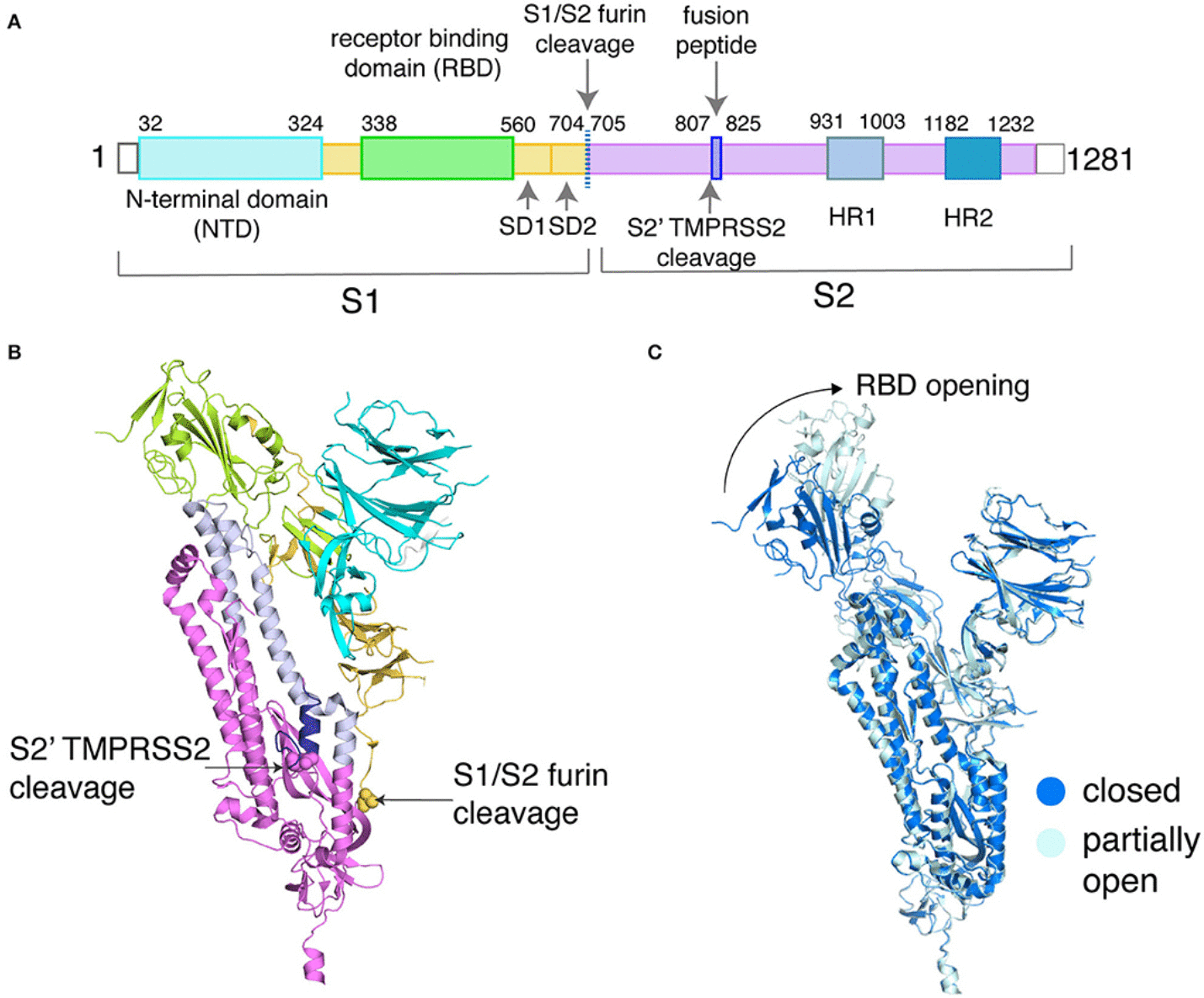}
    \caption{{\bf Sequence and structural organization of the SARS-CoV-2 spike protein:} (A) Domains of SARS-CoV-2 spike protein are depicted along with the two critical sites of cleavage. (B) These domains are shown on the structure of the spike protein (PDB-ID: 6VXX, only chain A is shown for clarity). The backbone is represented as a cartoon, and the domains are color-coded based on (A). The first residue at the two cleavage sites is highlighted as spheres and labeled. (C) Backbone alignment (chain B only) of the closed and partially open states of the spike protein reveals conformational changes at the RBD (shown by the arrow), with the rest of the domains showing RMSD $<$0.5 angstroms (a low backbone RMSD of $<$0.5 angstroms is also observed for chain A/C between the closed and partially open states). Figure reproduced from Reference \cite{DiPaola2021} (Creative Commons Attribution License (CC BY)).}
    \label{S1 Fig}
\end{figure}

\begin{figure}
    \centering
    \includegraphics[width=\textwidth]{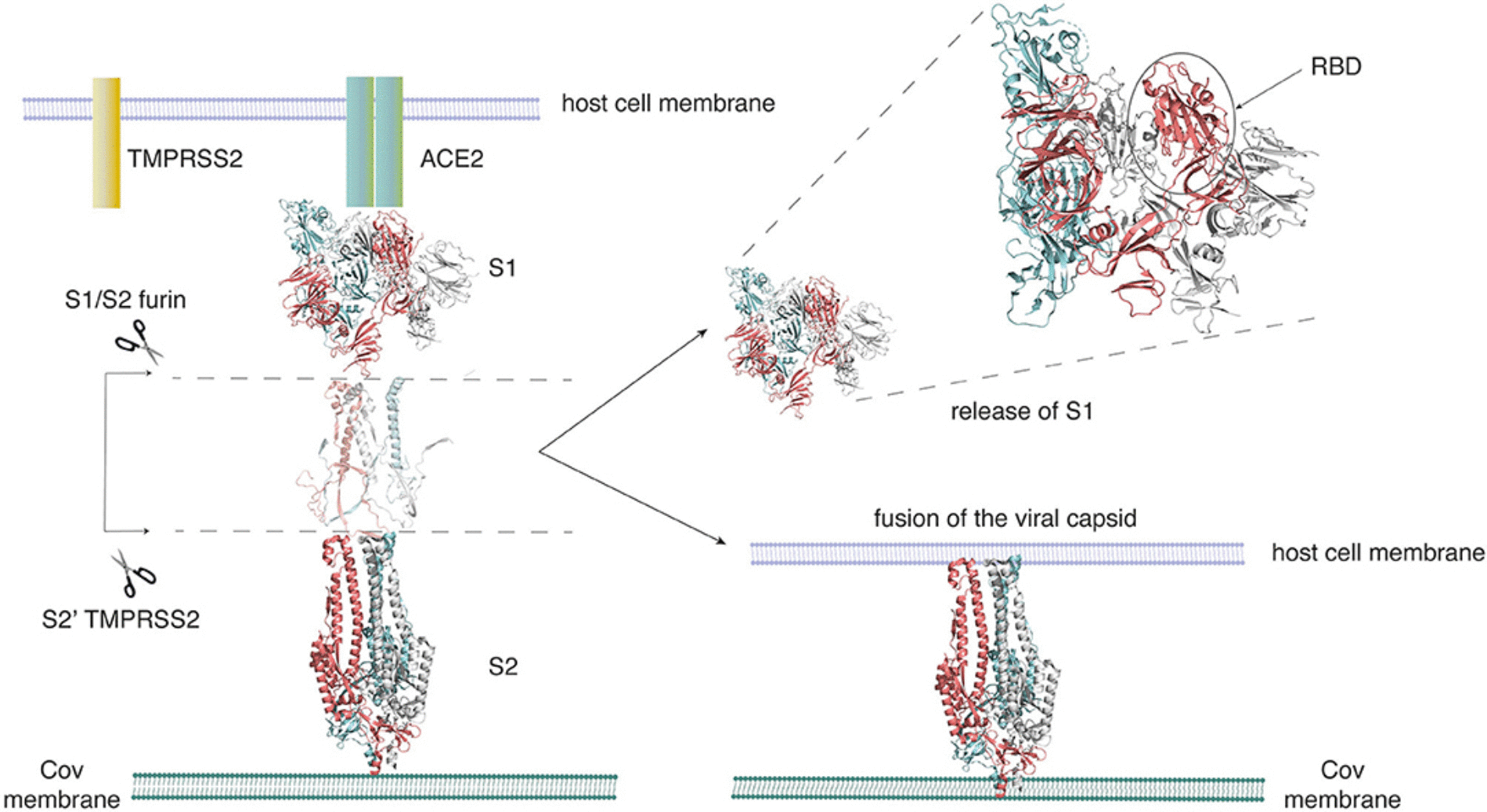}
    \caption{{\bf SARS-CoV-2 employs the spike glycoprotein to enter its host cell:}  The spike protein is composed of two domains, the S1 domain that hosts the receptor-binding domain (RBD), and the S2 domain. The S2 domain arbitrates the fusion of the viral and host cell membranes. Activation of the spike protein happens by cleavage at two sites (S1/S2 and S2\') by the Furin TMPRSS2 protease. The spike protein initially binds to the ACE2 receptor on the host cell through its RBD. On activation, it sheds the S1 domain, enabling S2 to fuse to the host cell membrane. This figure was adapted from Structural and functional mechanism of SARS-CoV-2 | Abcam (https://www.abcam.com/content/structural-and-functional-mechanism-of-sars-cov-2-cell-entry). Figure reproduced from Reference \cite{DiPaola2021} (Creative Commons Attribution License (CC BY)).}
    \label{S2 Fig}
\end{figure}

\begin{figure}
    \centering
    \includegraphics[width=\textwidth]{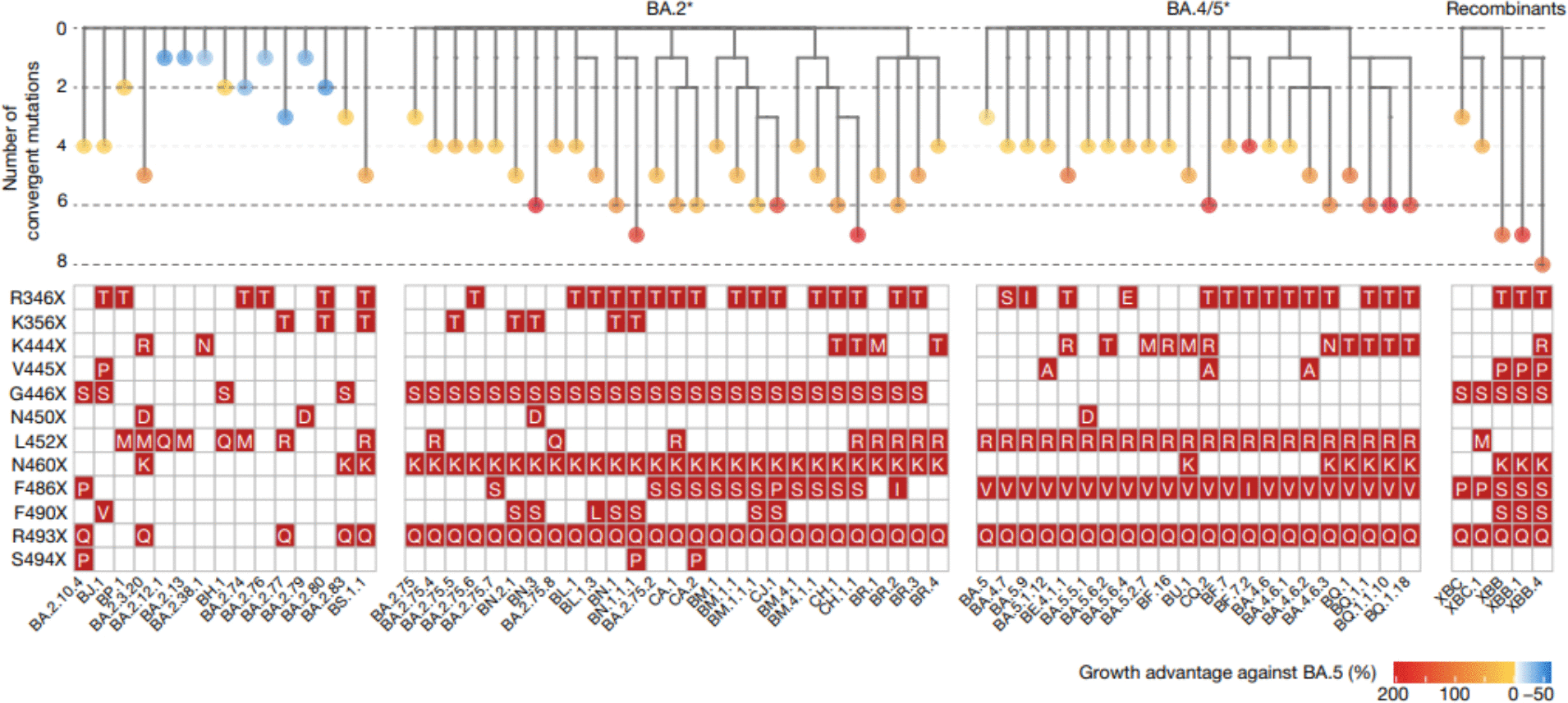}
    \caption{{\bf Post Omicron evolutionary path of SARS-CoV2 lineages leading to complex-recombinant strains: } This figure is reproduced as is from Reference \cite{Cao2023} under a Creative Commons Attribution 4.0 International License (\url{http://creativecommons.org/licenses/by/4.0/}).}
    \label{S3 Fig}
\end{figure}

\begin{figure}
    \centering
    \includegraphics[width=0.5\textwidth]{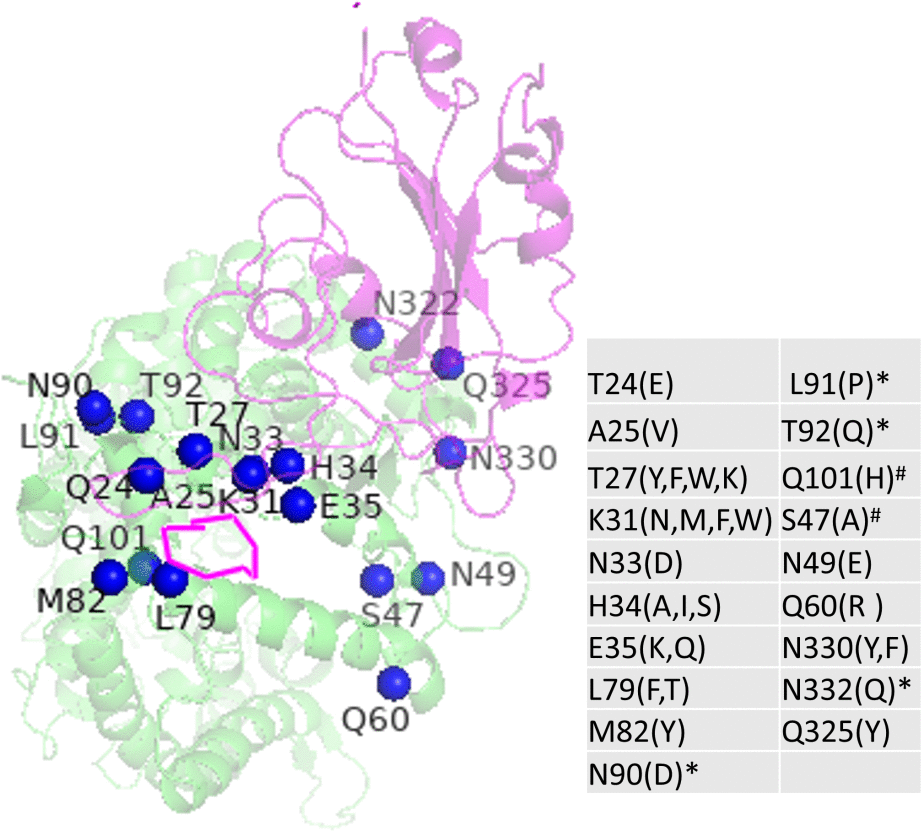}
    \caption{{\bf Mutation Sites for enhanced affinity in designing decoy ACE2 Receptors:} Possible residue substitutions are listed in the table. The loop held by the disulphide bridge in SARS-CoV2 (Cys480-Cys488) is also highlighted as a pink ribbon. This figure is loosely adapted from Reference \cite{Arimori2022-ls}.}
    \label{S4 Fig}
\end{figure}

\begin{figure}
    \centering
    \includegraphics[width=0.5\textwidth]{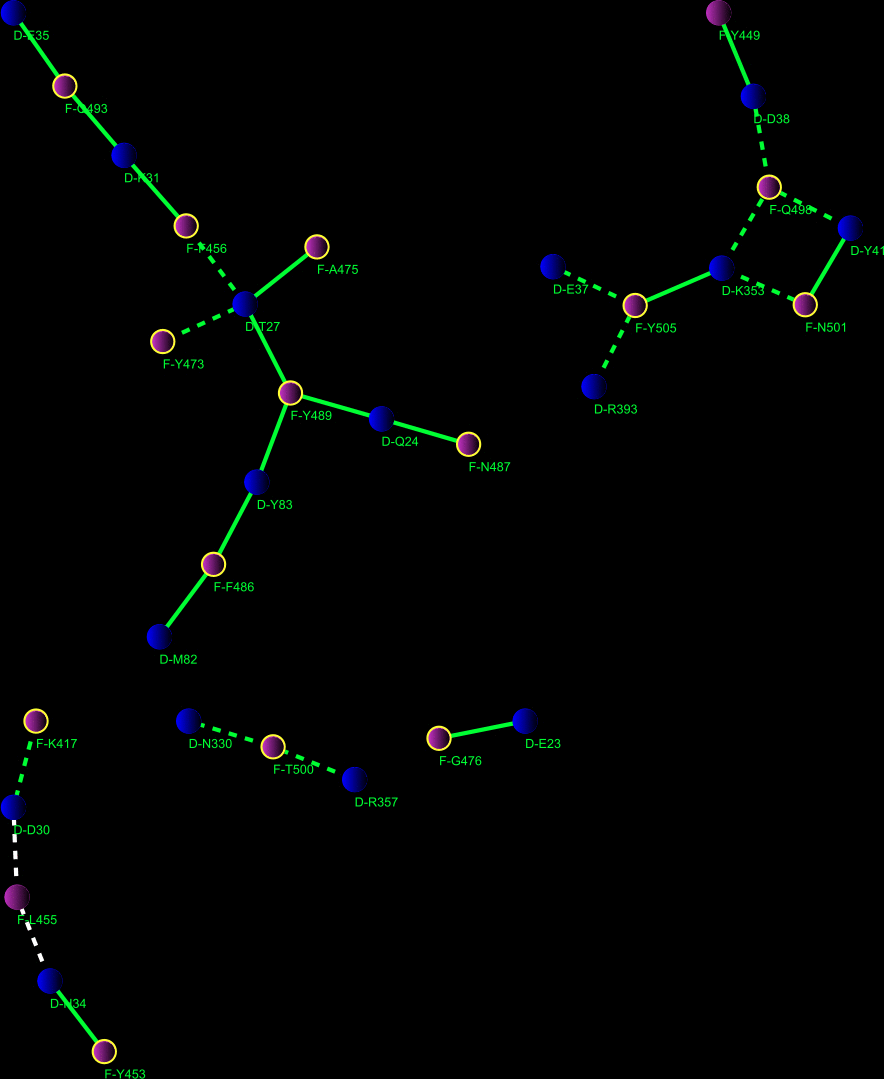}
    \caption{{\bf The Interface backbone of SARS-CoV2:} ACE2 receptor complex for the snapshot High RMSD (extracted at $I_{min}$=3.5). The pink and the blue spheres correspond respectively to the residues of SARS-CoV2 and the ACE2 receptor. The
green interface edge details are the same as given in Fig \ref{Fig 8}.}
    \label{S5 Fig}
\end{figure}

\begin{figure}
    \centering
    \includegraphics[width=0.5\textwidth]{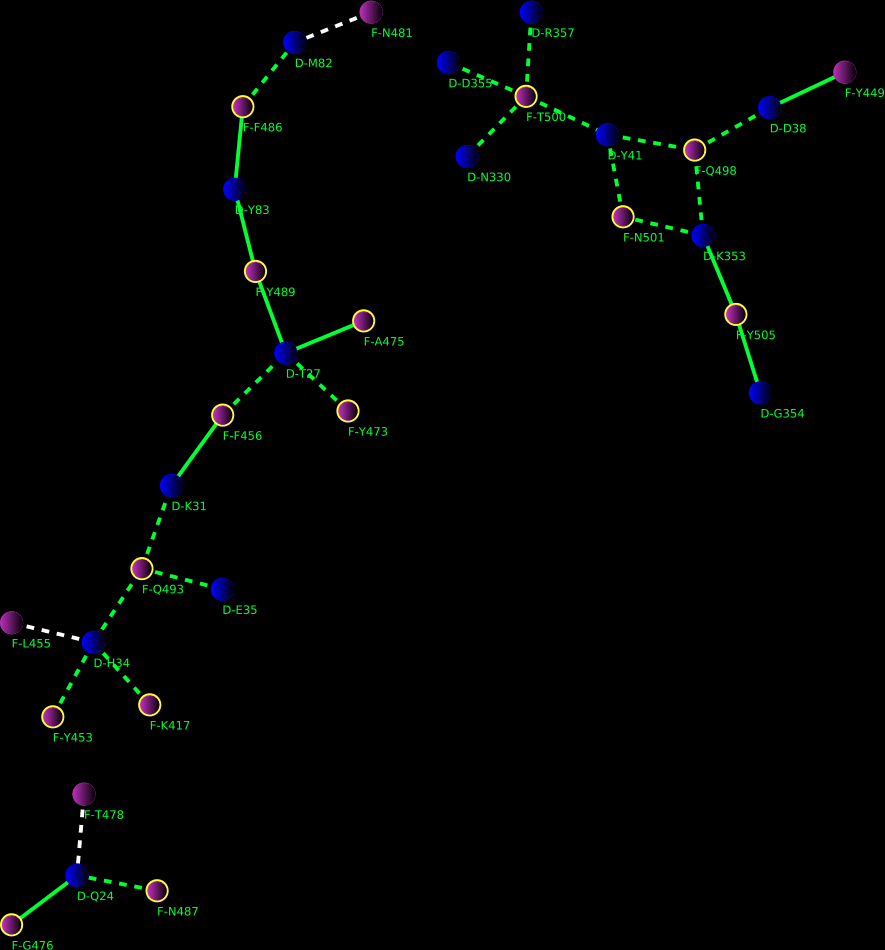}
    \caption{{\bf The Interface backbone of SARS-CoV2:} ACE2 receptor complex for the snapshot END Structure (extracted at $I_{min}$=3.5). The pink and the blue spheres correspond respectively to the residues of
SARS-CoV2 and the ACE2 receptor. The green interface edge details are
the same as given in Fig \ref{Fig 8}.}
    \label{S6 Fig}
\end{figure}

\end{document}